%% file: main.tex
\documentclass[10pt]{iopart}
\input{packages_commands}
\usepackage{esint}
\usepackage{amsmath}

\hypersetup{
     colorlinks   = true,
     linkcolor    = blue,
     citecolor    = blue,
     urlcolor     = blue
}

\begin{document}
\title{2D electron density profile evolution during detachment in Super-X divertor L-mode discharges on MAST-U }
\author{N. Lonigro$^{1,2}$, R. S. Doyle$^{2,3}$, K.Verhaegh$^4$, B. Lipschultz$^1$, D. Moulton $^2$, P. Ryan$^2$, J. S. Allcock$^2$, C. Bowman$^2$, D. Brida$^5$, J. Harrison$^2$, S. Silburn$^2$,C. Theiler$^6$,  T.A. Wijkamp$^{7}$, the EUROfusion Tokamak Exploitation Team$^{8}$ and MAST-U Team$^{9}$}
\address{$^1$York Plasma Institute, University of York, United Kingdom  }
\address{$^2$UKAEA, Culham Campus, Abingdon, Oxfordshire,OX14 3DB, United Kingdom }
\address{$^3$NCPST, Dublin City University, Ireland}
\address{$^4$Department of Applied Physics, Eindhoven University of Technology, Netherlands}
\address{$^5$ Max Planck Institute for Plasma Physics, Garching, Germany}  
\address{$^6$ Ecole Polytechnique Federale de Lausanne (EPFL), Swiss Plasma Center (SPC), Switzerland}
\address{$^7$DIFFER, Eindhoven, Netherlands}
\address{$^8$See the author list of E. Joffrin et al. 2024 Nucl. Fusion 64 112019}
\address{$^9$See the author list of J.R. Harrison et al 2024 Nucl. Fusion 64 112017}
\ead{\mailto{nicola.lonigro@ukaea.uk}}

\begin{abstract}
The improved performance of the long-legged alternative divertor configurations on MAST-U, combining strong baffling and total flux expansion, results in stronger power and momentum losses. These lead to lower electron temperatures and peak particle fluxes at the target compared to conventional divertor configurations. The evolution of the divertor electron density profile in the detached region is characterized by a density peak building up near the target, downstream of the detachment front, with increasing upstream density which then moves upstream from the target as the target temperature decreases ($T_e < 0.3$ eV). This behaviour is in agreement with simplified modeling based on the competition of different ion recombination processes and neutral drag acting on the plasma flow. Comparisons against SOLPS simulations generally show good agreement in the magnitude of the electron density profiles, both along and across the separatrix, but with some discrepancies in the shape of the profiles in the private flux region and near the target.
\end{abstract}


\section{Introduction}
Future tokamak fusion reactors are designed to generate a large amount of power. If unmitigated, a significant fraction of this power would be funneled to the divertor and concentrated in a thin toroidal ring when reaching the walls of the device, with expected poloidal widths on the order of millimiters \cite{DEMO_lambda}, and the associated heat fluxes on the divertor targets would surpass material limits. Thus, reactors will need to operate in detached divertor conditions to ensure enough volumetric particle, power and momentum losses occur to reduce the peak heat flux on the target through spreading the heat flux poloidally over larger regions of the divertor surface\cite{Wenninger_2014_DEMO_detach}. Easier access to detachment is favored, as it allows operation at lower core densities (and thus improved energy confinement\cite{core_confinement_density_degrad}) or allows sustaining higher heat fluxes entering the scape-off layer for the same heat flux reaching the target. A wider parameter space in which the divertor is stably detached is also positive, as it can act as a buffer and prevent the divertor from re-attaching during fuelling of heating transients\cite{kool2024demonstrationsuperxdivertorexhaust}. Understanding the detachment properties of a divertor, and being able to correctly model them, is then imperative to correctly extrapolate the behaviours of current experiments to future reactors. 

Alternative divertor configurations (ADCs) are being studied to determine whether these properties are improved compared to conventional designs. Examples of ADCs include the X-divertor\cite{Kotschenreuther2004ScrapeOL}, the Snowflake divertor\cite{Ryutov2007}, the Super-X divertor\cite{Valanju_2009} and the X-point target divertor\cite{LaBombard_2015}. Devices with flexible divertor magnetic topologies, such as TCV\cite{Piras_2009}\cite{maurizio_TCV_2018} and MAST-U \cite{soukhanovskii_mastu_2022},\cite{Verhaegh_2023}\cite{Verhaegh_2023_2}, are particularly useful in comparing many different divertor configurations with similar core plasma conditions.  The MAST-U Super-X divertor combines the advantages of large total flux expansion ($B_{Xpt}$$/$$B_{Target}$)\cite{Valanju_2009} with a long poloidal divertor leg length through the strongly baffled divertor chambers, increasing dissipative plasma-neutral interactions. The results of divertor plasma research on MAST-U have shown a strong improvement in exhaust performance when compared to more conventional divertors\cite{verhaegh2024improved}, in accordance with modelling\cite{moulton_super-x_2024}\cite{MAURIZIO2024101736}. These improvements include improved access to detachment, reduced sensitivity of the detachment front location and additional ion sinks \cite{Verhaegh_2023_2}. Measurements of increased molecular rotational temperature during detachment have been attributed to ion-molecule collisions \cite{Osborne_2024}, which also lead to plasma power and momentum losses. Reduced modelling also suggests that the longer leg may improve the ability to buffer fast transients through the additional volume of neutrals \cite{HENDERSON_reattachment}. Further work was dedicated to core-edge integration, demonstrating high core performance with a strongly detached divertor\cite{Harrison_2024}, compared to the conventional divertor in which the particle flux roll-over is associated with a degradation of the confinement time due to the high upstream density required\cite{FEDERICI2025101940}. Experiments on the Super-X divertor on TCV, instead, have shown no significantly improved detachment characteristics in a non-baffled configuration \cite{Theiler_2017} and only modest improvements in a baffled configuration when compared to simple analytical scalings\cite{Theiler_2024}. Possible explanations for part of this discrepancy are the effect of strong plasma flows on the benefits of flux expansion (and the related momentum losses)\cite{Carpita_2024} and the difference in baffling between the two machines\cite{fil_TCV_2020}. Improved understanding on the basic physic processes related to detachment and their consequences on the plasma parameters profiles can help in better understanding these discrepancies.

In this work, the 2D electron density profiles obtained with the new multi-delay coherence imaging (CIS) diagnostic \cite{Doyle_CIS}\cite{lonigro_CIS} are studied in a variety of divertor magnetic topologies under similar upstream conditions and compared against models to determine the influence of plasma shaping on the divertor electron density behaviour and its evolution in deeply detached conditions in a strongly baffled divertor. These measurements allow validating high-fidelity modelling tools, such as SOLPS simulations, and can be a base to build reduced models which capture the main physics dominating the divertor behavior.  In section \ref{sec:MAST-U}, an overview of the MAST-U device is given and the divertor configurations of interest for this paper are shown. In section \ref{sec:measurements} the 2D profile measurements are compared in a variety of L-mode conditions. The density downstream of the detachment front is inferred to increase with increasing upstream density, before peaking and starting to decrease in the deepest stages of detachment. The target density is inferred to decrease with increasing leg length and total flux expansion for similar upstream conditions. These measurements are discussed and compared to reduced models and SOLPS simulations in section \ref{sec:discussion}, suggesting an increase in power and momentum losses with increasing strike point radius and that the evolution of the density profile downstream of the detachment front may be described as a competition of neutral collisions slowing down the plasma and recombination ion sinks.

\section{MAST-U discharges overview}
\label{sec:MAST-U}
\begin{figure}[]\centering
\includegraphics[width= 0.58\textwidth]{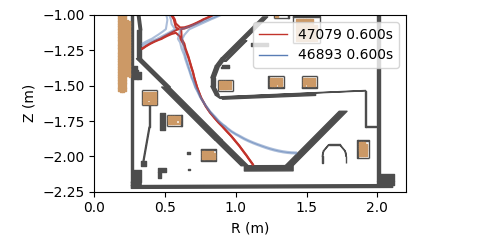}
\vspace{-0.4 cm}\caption{Range of divertor configurations scanned in this work. Taken from EFIT magnetic reconstructions at t = 0.6 s of the core density ramps in elongated divertor \#47079 and Super-X divertor \#46893.}
\label{fig:fig1_mag_recon}
\vspace{-0.3cm} 
\end{figure}

\begin{figure}[]\centering
\includegraphics[width= 0.5\textwidth]{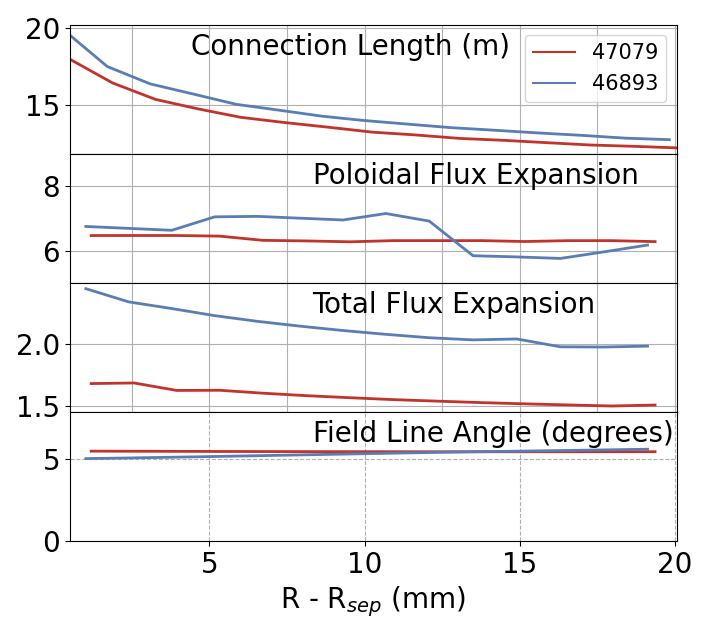}
\vspace{-0.2 cm}\caption{Divertor geometry parameters at the target in ED(\#47079) and SXD(\#46893), plotted as function of distance from the separatrix at the outer midplane.}
\label{fig:fig3_ramps_div}
\vspace{-0.3cm} 
\end{figure}

MAST-U is a medium sized spherical tokamak with major radius R = 0.9 m and minor radius a = 0.6 m, thus with a small aspect ratio of A (R/a) = 1.5. All the discharges in this work have a toroidal field on axis of $\sim$ 0.65 T and a plasma current of $\sim$ 750 kA. Gas fuelling is performed in the main chamber from the low field side to inhibit access to H mode\cite{LFS_H_mode_access}. Working in L mode prevents ELMs, thus simplifying the study of the density profile evolution using the CIS diagnostic, which in the 2nd MAST-U campaign operated with a framerate of 30 Hz. The  divertor leg is scanned from the elongated divertor (ED)\cite{verhaegh2024improved} magnetic configuration to the Super-X (SXD) divertor configuration, both in a symmetric double-null configuration, as shown in figure \ref{fig:fig1_mag_recon} and described further in previous work \cite{verhaegh2024improved}. The focus here is on long-legged configurations due to the better diagnostic coverage of these geoemetries, compared to the conventional divertor.  
Some details on the divertor magnetic geometries are given in figure \ref{fig:fig3_ramps_div}, where the connection length, poloidal flux expansion $(F_X = B_\theta^MB_\phi^T/B_\theta^MB_\phi^T$, with $M$ and $T$ denoting the midplane and target while $\phi$ and $\theta$ denoting the toroidal and poloidal field components), total flux expansion $(F_R = B^{X_{pt}}/B^T)$ and total field line angle at the target, thus including both poloidal tilt of the tile and toroidal tilt of the field line, are compared. Additional information on the different geometries can be found in \cite{verhaegh2024improved}.
The main difference is the increase in total flux expansion in the SXD compared to the ED, although the SXD also has a longer connection length, both parallel to the total magnetic field and in the poloidal plane, as well as a larger poloidal flux expansion. The strike point field line angle with the target, which can impact the heat flux footprint on the target and affect the direction of the recycled neutrals with respect to the separatrix\cite{Loarte_2001_review}, is comparable between the two geometries.
\begin{figure}[]\centering
\includegraphics[width= 0.5\textwidth,trim={1.2cm 0 0 0},clip]{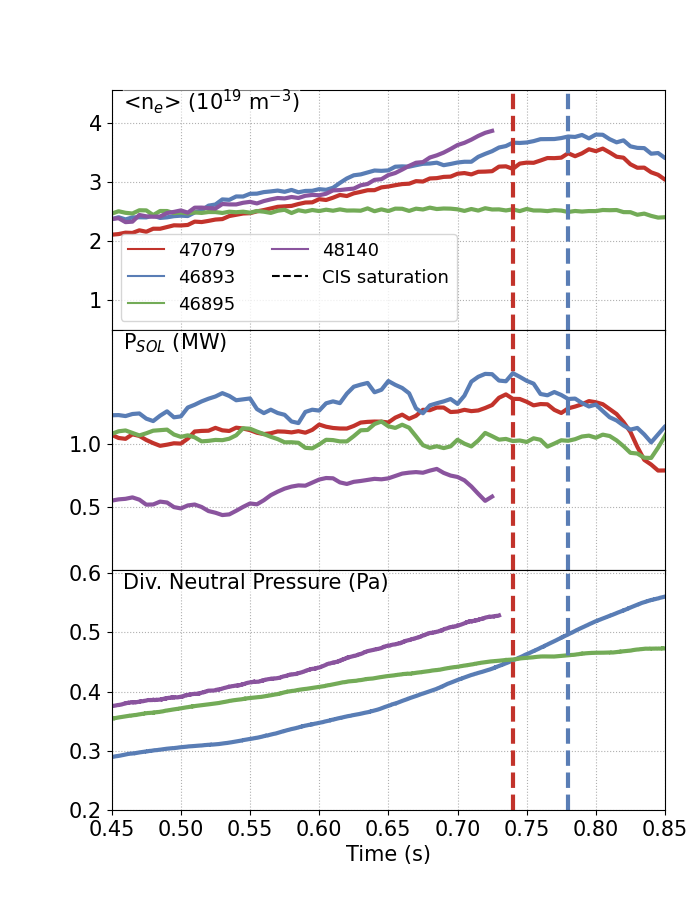}
\vspace{-0.1 cm}\caption{Overview of core parameters (line-averaged density, $P_{SOL}$) and divertor pressure in the discharges analyzed in this work: NBI core density ramps in SXD (\#46893) and ED (\#47079), NBI heated strike point radius scan (\#46895) and Ohmic core density ramp in SXD (\#48140). The vertical dashed lines show the time at which the CIS camera saturates in discharges \#47079 and \#46893. No divertor pressure measurements are available in \#47079.}
\label{fig:fig2_ramps_ov}
\vspace{-0.3cm} 
\end{figure}
NBI-heated core density ramps (P$_{NBI}\sim$  1.5 MW, P$_{SOL}\sim$
1 - 1.2 MW \footnote{Estimated experimentally as $P_{SOL} = P_{Ohmic} + P_{NBI} - P_{\dd W/ \dd t} - P_{rad}$, with $P_{Ohmic}, P_{NBI}, P_{\dd W/ \dd t} , P_{rad}$ the ohmic heating power, the power absorbed from the NBI, the variation in internal energy and the core radiated power respectively.}) have been performed in both the ED and SXD magnetic configurations to study the evolution of the electron density profiles with increasing depths of detachment at the two extremes of the scanned target major radius range. An ohmic core density ramp ( P$_{SOL}\sim$ = 0.6 MW) in SXD is also analyzed to study more deeply detached conditions. The Elongated and Super-X divertor are never attached in the experimental range examined in this work. The minimum core density achievable in the core density ramps is set by the onset of MHD activity, which manifests itself in the divertor in the form of strike point splitting\cite{wijkamp_MWI_2023}. Such conditions are avoided as they can break toroidal symmetry( an assumption used in the camera analysis), and they complicate comparisons with higher density conditions.  The discharges discussed in this work are:
\begin{itemize}
    \item \#48140 Ohmic core density scan in Super-X divertor configuration
    \item \#46893 Beam-heated core density scan in Super-X divertor configuration
    \item \#47079 Beam-heated core density scan in Elongated divertor configuration
    \item \#46895 Beam-heated strike point radius scan with constant upstream conditions
\end{itemize}
An overview of the core parameter evolution during these shots is shown in figure \ref{fig:fig2_ramps_ov}.
The long-legged divertor shapes are established at 0.45 s, and the scanned range of the line-averaged core electron density is comparable in the different density ramps, 2-4 $\cdot 10^{19}m^{-3}$. This corresponds to a range of [25 \% , 50 \%] in Greenwald density fraction. The core electron density is kept constant during the divertor leg scan, albeit the divertor neutral pressure keeps increasing during the discharge. The NBI power is slightly higher in the SXD core density ramp, 1.7 MW, compared to the 1.5 MW in the ED core density ramp and the shape scan, resulting in a slightly higher $P_{SOL}$. The increasing trend of $P_{SOL}$ during the density ramps is attributed to an increased fraction of the NBI power being absorbed at higher density. The Ohmic discharge has no additional NBI power and thus a lower $P_{SOL}$.
\section{Electron density profiles measurements}
\label{sec:measurements}
MAST-U combines state of the art diagnostic techniques, such as multi-delay coherence imaging\cite{Allcock_CIS_density}\cite{Doyle_CIS} and multi-wavelength imaging\cite{feng_development_2021}\cite{wijkamp_MWI_2023}, with a long divertor poloidal leg in a baffled divertor chamber. The strong baffling results in strong neutral trapping, with the majority of the divertor ion source being generated downstream of the X-point\cite{verhaegh2024improved}. The divertor chamber allows a tangential view of the divertor leg and can make it easier to diagnose compared to more compact baffled designs, while the long divertor leg reduces requirements on spatial resolution and can more easily allow detailed studies of the divertor behaviour along the separatrix. \\
The electron density is one of the basic parameters describing the divertor state, with most of the processes driving detachment, power and momentum losses having electron density dependencies (e.g. three-body recombination scaling as $n_e^3$). This work will focus on characterizing the evolution of this fundamental parameter, before moving on to the evolution of derived quantities in future work. In particular, there is a focus on the behaviour of the electron density in detached conditions, which will be required for the safe operation of future devices.\\
The CIS diagnostic uses interferometric imaging techniques to obtain Stark broadening information of the deuterium (D) Balmer gamma spectral line (5 $\rightarrow$ 2, 434 nm), which is directly related to the electron density. This information is then used to infer 2D poloidal profiles under the assumption of toroidal symmetry, enabling detailed studies on the electron density profile evolution. Details on the inference technique, its validation on synthetic data, and comparison against other diagnostics have been previously reported \cite{lonigro_CIS}. 
\subsection{Evolution during detachment}\label{sec:evol_during_detach}
\begin{figure*}[]\centering
\begin{subfigure}[b]{0.49\textwidth}
\includegraphics[width=\textwidth]{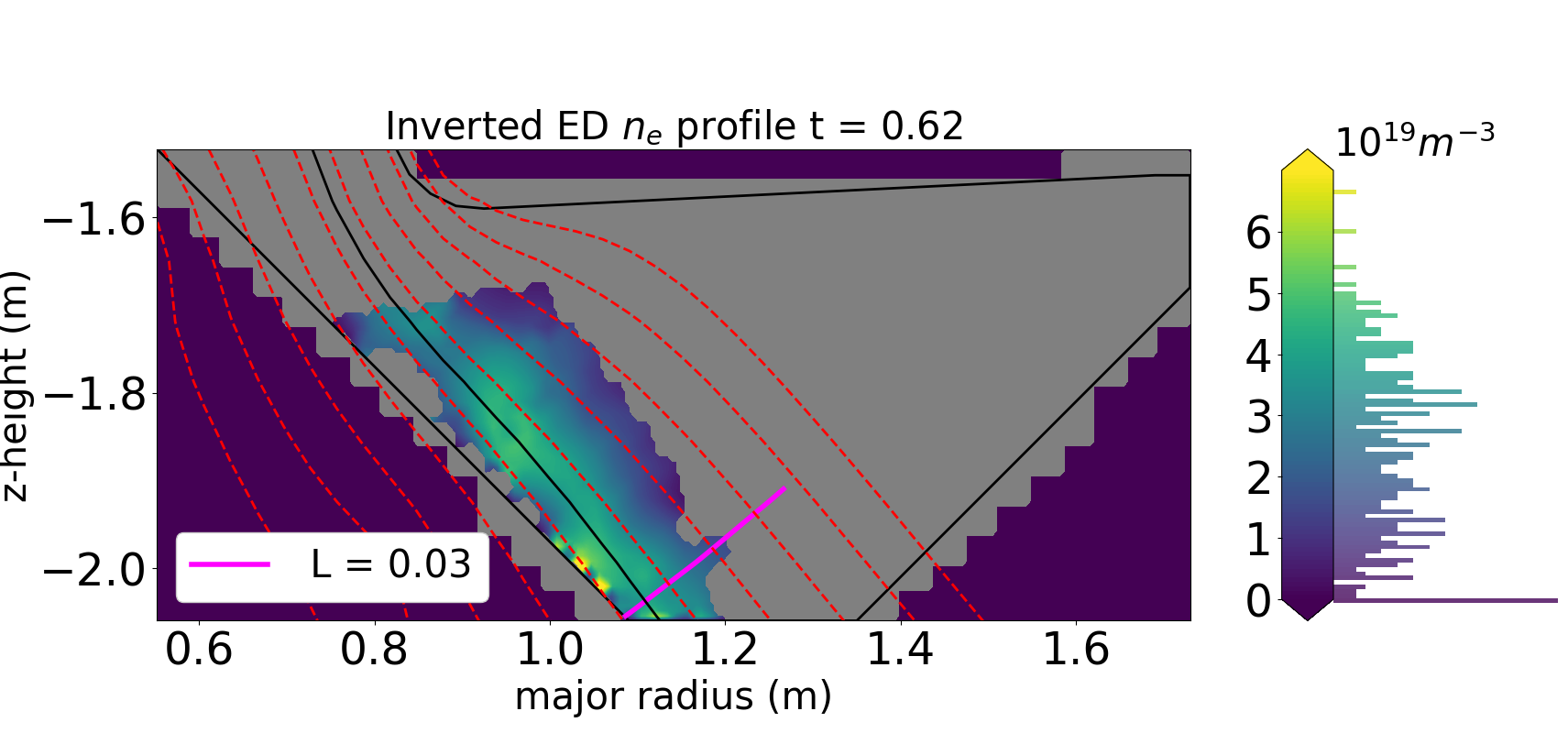}
\caption{}
\end{subfigure}
\begin{subfigure}[b]{0.49\textwidth}
\includegraphics[width=\textwidth]{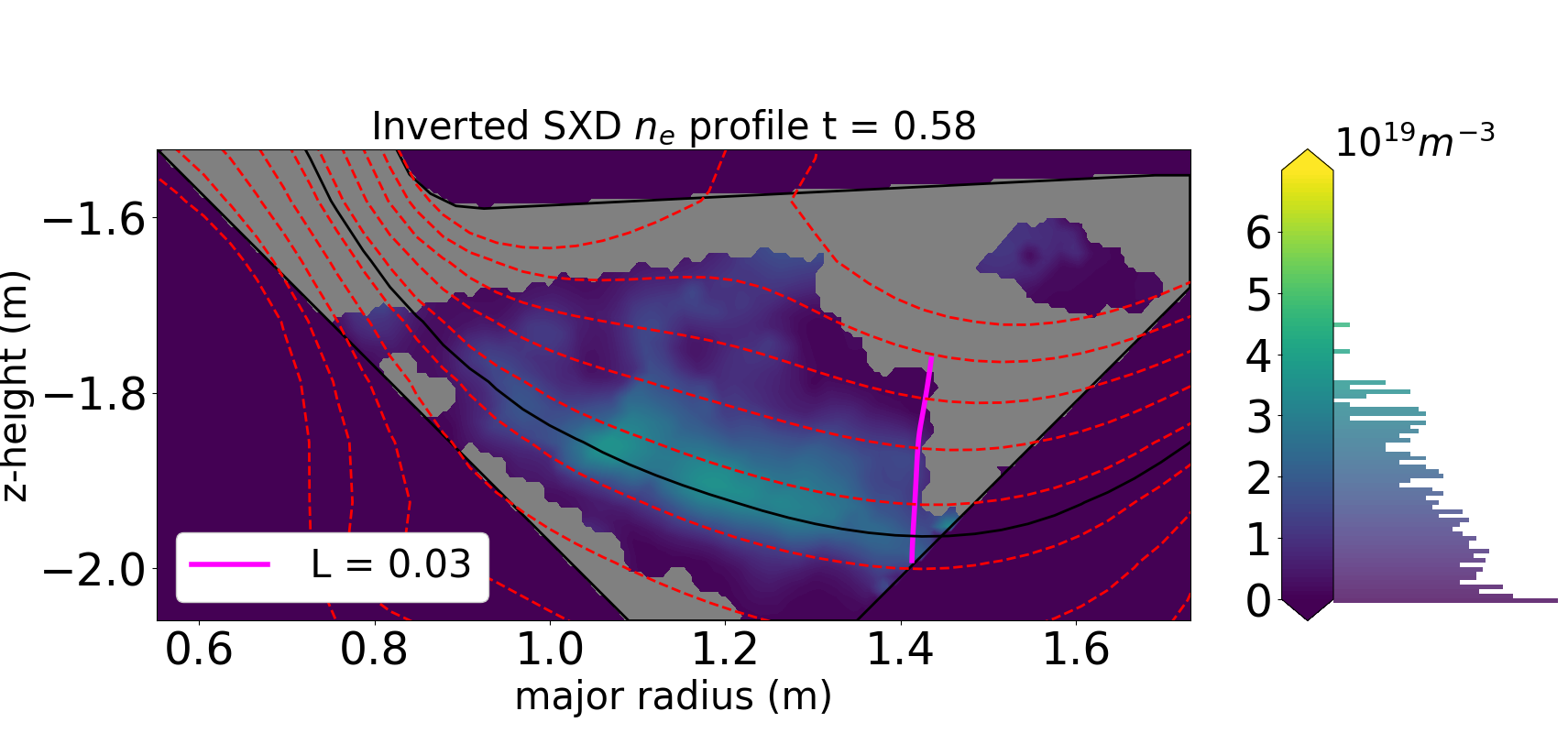}
\caption{}
\end{subfigure}
\vspace{-0.3 cm}\caption{2D electron density profile in NBI heated (a) ED core density ramp \#47079 and (b) SXD core density ramp \#46893 at 40 \% Greenwald fraction. The separatrix is shown in black. The masked off region has an emissivity $\leq$ 10 \% of its maximum. A histogram of the plotted values is shown along the colorbar. The cross-field line starting at 3 cm of poloidal distance from the target (L = 3 cm) along the separatrix, studied further in figure \ref{fig:fig7_cross_profiles}, is also shown.}
\label{fig4:2D_profiles}
\vspace{-0.3cm} 
\end{figure*}

The behaviour of the divertor density profiles with increased depth of detachment can be compared at the two extremes of the scanned divertor leg strike point range using core density ramps with fixed divertor shapes and similar core conditions. The SXD and ED divertor plasmas are already detached at the start of the discharge, but as the core density increases the divertor plasmas evolve as the target electron temperature decreases further \cite{verhaegh2024improved}. Towards the end of the discharge, as the plasma emission increases significantly due to the onset of electron-ion recombination emission, the CIS camera saturates. Thus, the density profiles could only be reconstructed for a part of the density scans, as shown in figure \ref{fig:fig2_ramps_ov} before the dashed vertical bars. A comparison of the 2D electron density profiles for the same core Greenwald fraction of 40\% are shown in figure \ref{fig4:2D_profiles}.
The electron density is higher in the ED configuration compared to the SXD; the poloidal spatial profile is mostly flat in both configurations. 

\begin{figure}[]\centering
\begin{subfigure}[b]{0.45\textwidth}
\includegraphics[width=\textwidth]{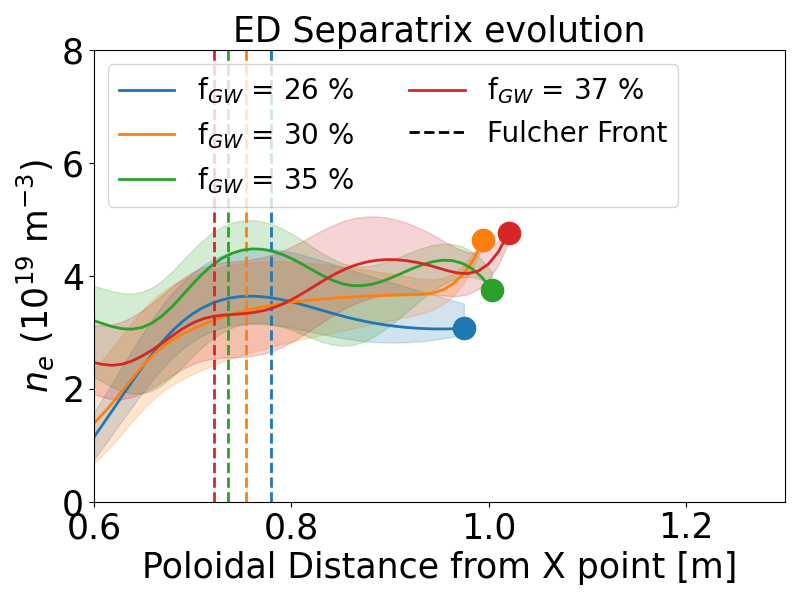}
\caption{}
\end{subfigure}
\begin{subfigure}[b]{0.45\textwidth}
\includegraphics[width=\textwidth]{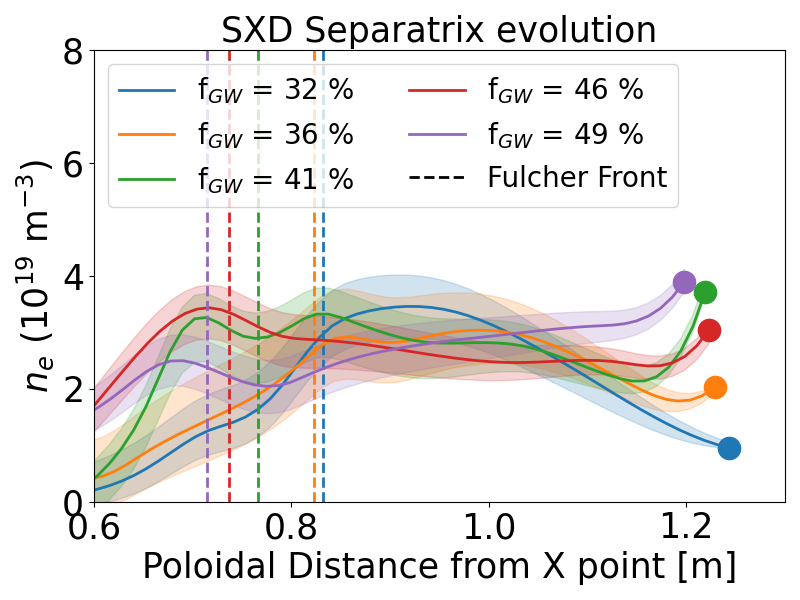}
\caption{}
\end{subfigure}
\vspace{-0.3 cm}\caption{Separatrix electron density evolution in NBI heated (a) ED core density ramp \#47079 and (b) SXD core density ramp \#46893, with the uncertainty in the profiles shows as a shaded region. The 50 \% falling edge of the Fulcher emission  along the separatrix is plotted as vertical lines. The circle markers show the position of the target.}
\label{fig:fig5_sep_evol}
\vspace{-0.3cm} 
\end{figure}

\begin{figure}[]\centering
\begin{subfigure}[b]{0.45\textwidth}
\includegraphics[width= \textwidth,trim={0cm 0cm 0cm 0cm},clip]{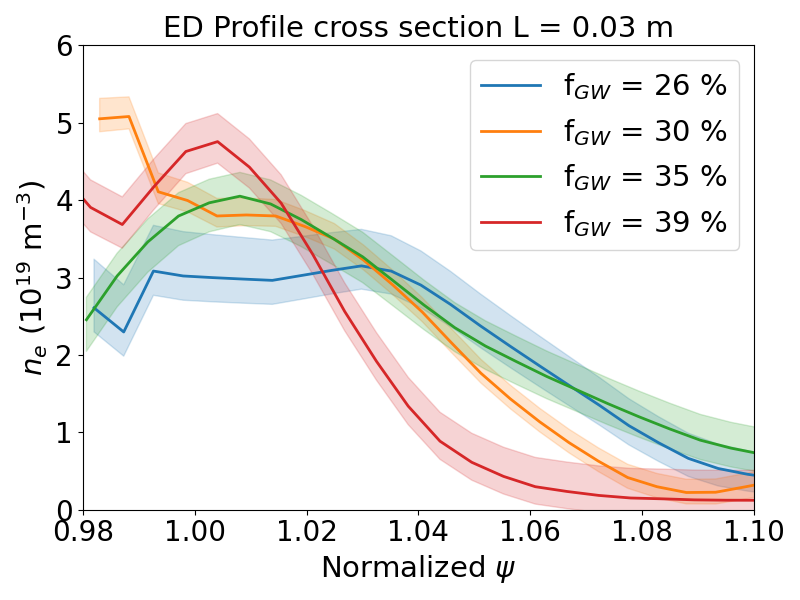}
\caption{}
\end{subfigure}
\begin{subfigure}[b]{0.45\textwidth}
\includegraphics[width= \textwidth]{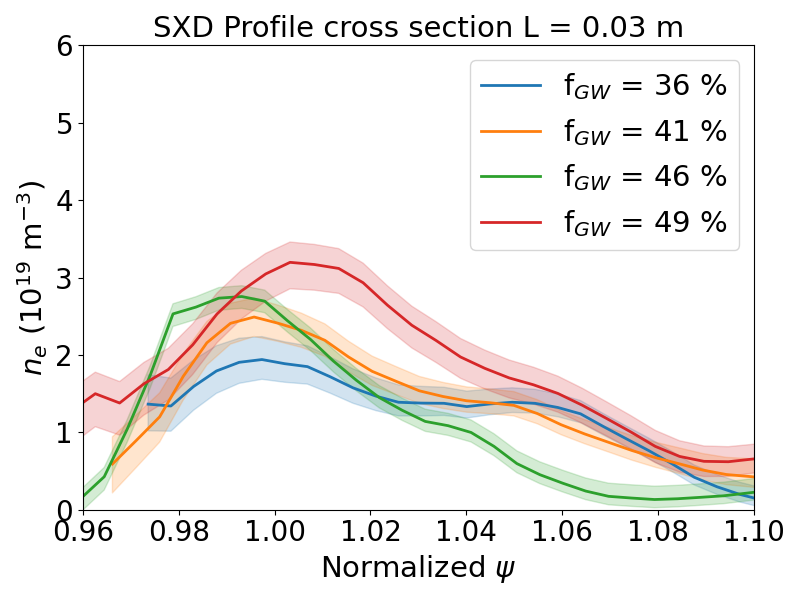}
\caption{}
\end{subfigure}
\vspace{-0.3 cm}\caption{Separatrix cross-field electron density profiles (a) during the NBI heated ED density scan \#47079 and (b) SXD density scan \#46893 along the cross-field section starting at 3 cm of poloidal distance from the target along the separatrix (shown in figure \ref{fig4:2D_profiles}), with the uncertainty in the profiles show as a shaded region.}
\label{fig:fig7_cross_profiles}
\vspace{-0.3cm} 
\end{figure}

The evolution of the density profiles during the density ramp along the separatrix determined by EFIT are shown in figure \ref{fig:fig5_sep_evol} for the two discharges, along with the position of the 50 \% falling edge ("emisison front") for the $D_2$ Fulcher band emission along the separatrix, used as a proxy for the detachment front location \cite{Verhaegh_2023}\cite{Verhaegh_2023_2}. This allows studying the evolution of the separatrix density profiles with increasing core density, expressed as a percentage of the Greenwald fraction $f_{GW}$, as the divertor becomes more deeply detached.
The $n_e$ spatial profile along the separatrix is flat for most of the divertor chamber in both configurations, while significant density changes are observed close to the target: at the lowest core densities, the density profiles are non-monotonic, slightly decreasing between the detachment front and the target. As the core density increases, the spatial profile flattens and then starts peaking towards the target at even higher core densities. The density peak is not observed to detach from the target even in the most detached conditions accessed in this discharge, although this is observed in discharges with lower $P_{SOL}$, as discussed in section \ref{sec:dens_detach}.
In figure \ref{fig:fig7_cross_profiles}, the evolution of the cross-field electron density profiles during the ED and SXD scans are mapped to the midplane and shown in normalized flux coordinate space ($\bar{\psi}$)\footnote{ In the divertor the separatrix has a normalized flux coordinate $\left(\bar{\psi}  = \frac{\psi-\psi_{axis}}{\psi_{edge}-\psi_{axis}}\right)$ of 1, separating the private flux region ($\bar{\psi}<$1) and the SOL ($\bar{\psi}>$1)} at 3 cm from the target (L = 0.03 m) along the separatrix, to avoid possible inversion artifacts that can develop close to the target tiles \cite{lonigro_CIS}.
As the core and target density increase, the spatial profile in the ED configuration becomes more peaked $(\bar{\psi} \sim 1.01)$. A high electron density is inferred in the private flux region, which may be attributed to the proximity of the divertor leg to the neighboring tile, but it is unclear if this is a real effect or an artifact of the inversion, as also discussed in section \ref{sec:SOLPS}. In the SXD, the profile initially becomes more peaked, but then starts broadening again towards the end of the discharge, possibly correlated to the approaching detachment of the electron density front from the target which is not reached in this discharge before the CIS camera saturates.

\begin{figure}[]\centering
\begin{subfigure}[b]{0.45\textwidth}
\includegraphics[width= \textwidth]{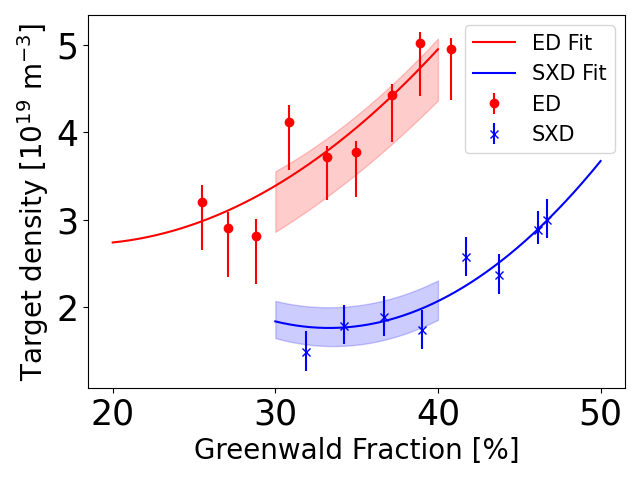}
\caption{}
\end{subfigure}
\begin{subfigure}[b]{0.45\textwidth}
\includegraphics[width= \textwidth]{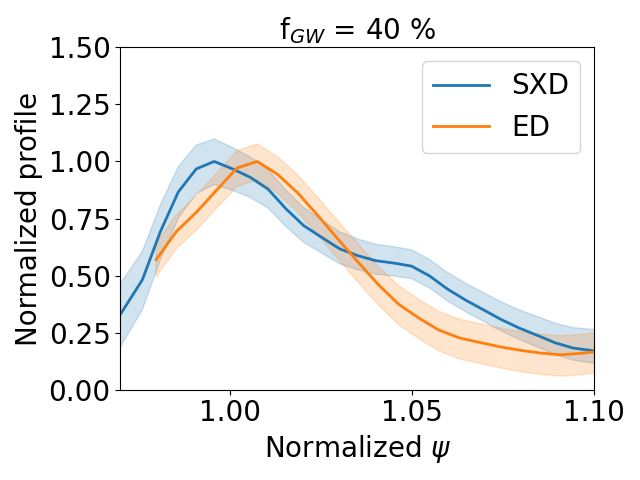}
\caption{}
\end{subfigure}
\vspace{-0.3 cm}\caption{(a) Evolution of separatrix electron density in NBI heated core density ramps in  ED (\#47079) and SXD (\#46893) configurations at 3 cm of poloidal distance from the target, fit with 2nd order polynomial function. (b) Normalized cross-field density profiles in the SXD and ED at the same core density (40 \% Greenwald fraction) starting at 3 cm of poloidal distance from the target along the separatrix.}
\label{fig:fig6_ramps_div}
\vspace{-0.3cm} 
\end{figure}

A direct comparison between the two shapes can also be made.
The density on the separatrix at 3cm from the target is taken as a proxy for the target density and it is plotted for both configurations as a function of upstream density in figure \ref{fig:fig6_ramps_div}. The shaded area highlights the region as a function of Greenwald fraction for which overlapping density data is available. A comparison of the  normalized cross-field profile in the ED and SXD configurations at the same upstream density (40 $\%$ Greenwald fraction) is also shown as a function of poloidal flux, normalizing the profiles by their maximum. 
The target density is consistently higher in ED configuration, with typical target densities up to $2 \cdot 10^{19}$ m$^{-3}$ larger than in the SXD. This could be attributed to the previously reported particle sinks present in the additional divertor volume available in the SXD \cite{verhaegh2024nbi}, lowering the electron density before the target is reached. Both ED and SXD target densities increase by $\sim$ 50 $\%$ as the upstream density increases by $\sim$ 50 \% during the ramp. Regions further away from the target, in the middle of the divertor chamber, do not show a strong variation in density throughout the discharge. Comparing the cross-field profiles in the two configurations shows a broader electron density profile in the SXD for the same upstream density even when plotting the profiles against poloidal flux and thus accounting for the expected broadening in real space due to poloidal flux expansion. This may suggest additional cross-field transport in the SXD compared to the ED, which could be attributed to the additional poloidal leg length if cross-field transport is constant along the divertor leg.
\subsection{Variation with strike point radius at constant core density}
\begin{figure}[]\centering
\begin{subfigure}[b]{0.45\textwidth}
\includegraphics[width= \textwidth]{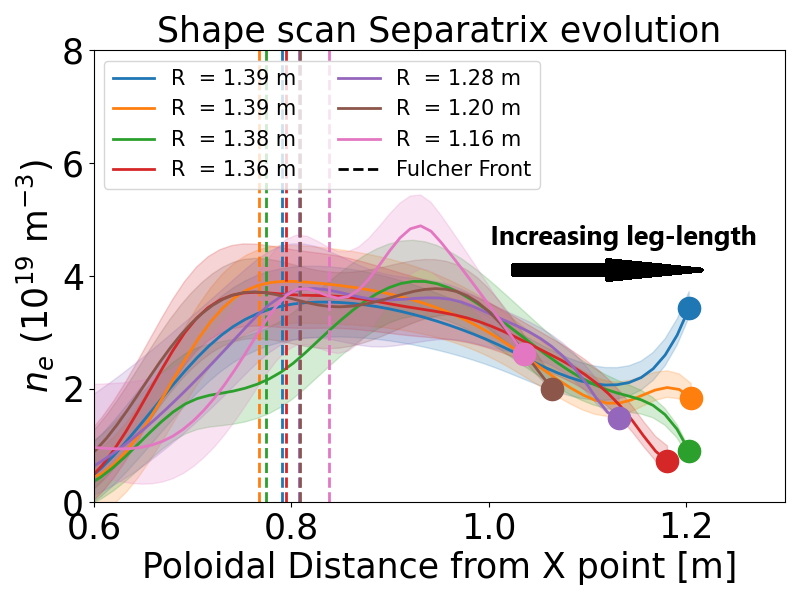}
\caption{}
\end{subfigure}
\begin{subfigure}[b]{0.45\textwidth}
\includegraphics[width= \textwidth]{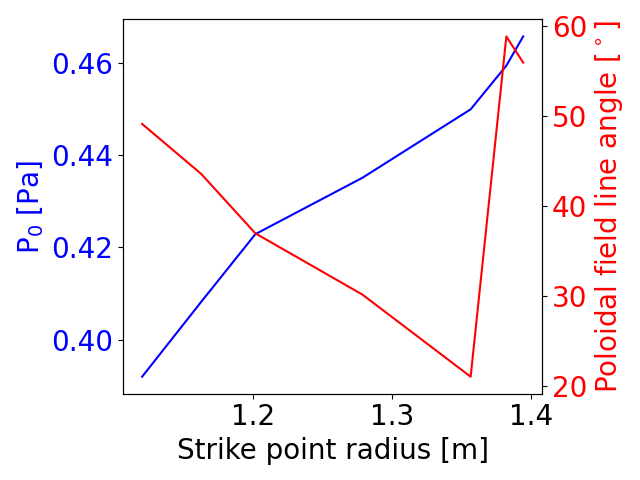}
\caption{}
\end{subfigure}
\vspace{-0.3 cm}\caption{(a) Separatrix density evolution in NBI heated poloidal leg length scan \#46895, with the uncertainty in the profiles shown as a shaded region.(b) Divertor neutral pressure measured by the pressure gauge and poloidal field angle between the separatrix and the tile as the strike point radius is increased.}
\label{fig:fig8_46895_cuts}
\vspace{-0.3cm} 
\end{figure}

\begin{figure}[]\centering
\includegraphics[width= 0.45\textwidth]{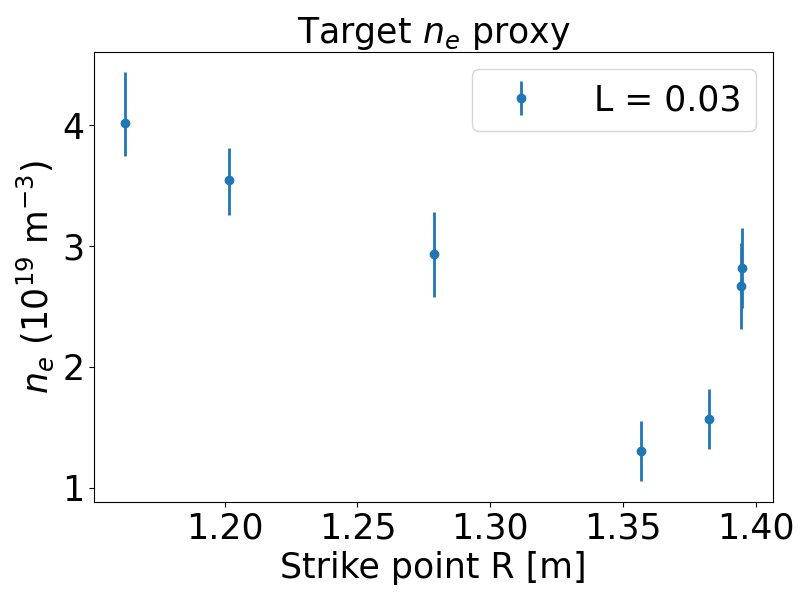}
\vspace{-0.3 cm}\caption{Target density in NBI heated strike point scan \#46895.}
\label{fig:fig9_46895_target}
\vspace{-0.3cm} 
\end{figure}

In discharge \#46895 the outer strike point radius is slowly swept from an Elongated to a Super-X configuration in L-mode to study the effect of changing magnetic geometry on the divertor behaviour for constant core parameters. The discharge has 1.5 MW of NBI power ($P_{SOL} \sim $ 1 MW) and a constant upstream density of $\sim$ 30 \% Greenwald fraction. The electron density profiles inferred by CIS as the leg moves from ED to SXD along the EFIT separatrix are shown in figure \ref{fig:fig8_46895_cuts}a as a function of distance from the X-point. The target position is shown with a dot marker.
The target electron density decreases with increasing strike point radius, albeit it appears to start rising again in the latest stages of the discharge. This could be possibly due to the continuously increasing divertor pressure or a change in poloidal angle between divertor leg and divertor tile as the leg moves between tiles, as shown in figure \ref{fig:fig8_46895_cuts}b. The particle flux to the target measured by Langmuir probes is monotonically decreasing, as will be discussed in section \ref{sec:analytical_models}. As in the core density ramps, the divertor plasma is always detached in the scanned strike point range. A non-monotonic electron density profile is inferred throughout the discharge, as observed in the initial stages of the core density ramps and in agreement with the low core density of this discharge. An interesting feature is that the density profiles appear to lay on top of each other when expressed in terms of poloidal distance from the X-point. This is in agreement with previous results showing that the Fulcher emission front (proxy for the $\sim$ 5 eV front position) remains in the same position when scanning the divertor leg\cite{verhaegh2024improved}, suggesting that (once it is off the target) the thermal front position is only a function of the magnetic geometry upstream of it, in agreement with simplified models \cite{Lipschultz_2016}\cite{Cowley_2022}. The behaviour of the poloidal density profiles inferred by CIS seems to suggest that something similar might apply for the electron density, likely due to similar particle loss (recombination) and source (ionization) terms as a function of distance from the X-point\cite{verhaegh2024improved}. In figure \ref{fig:fig9_46895_target} the evolution of the separatrix density at 3 cm from the target is shown.
As the leg sweeps out, the electron density near the target decreases linearly with strike point major radius. The higher target density at a major radius of 1.4 m is associated with the separatrix crossing between two tiles, possibly due to a change in the poloidal angle between the separatrix and the target tile affecting the direction of the recycling neutrals and resulting in stronger plasma-neutral interactions \cite{Loarte_2001_review}.
\subsection{Density front detachment in lower power conditions}\label{sec:dens_detach}
\begin{figure}[]\centering
\includegraphics[width= 0.55\textwidth]{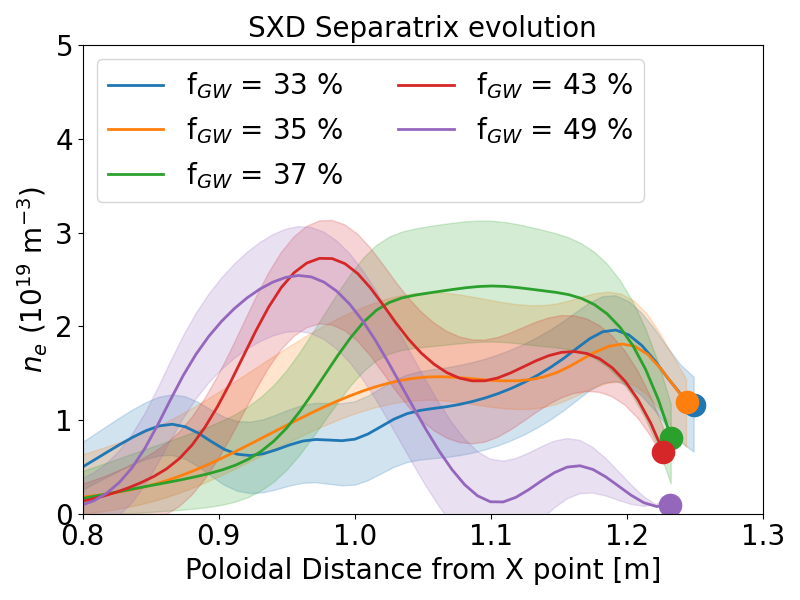}
\vspace{-0.3 cm}\caption{Separatrix density evolution during Ohmic core density ramp \#48140, with the uncertainty in the profiles shows as a shaded region. The circle markers represent the target position. The fulcher front is further upstream than the divertor leg section shown.}
\label{fig:fig10_48140_sep}
\vspace{-0.3cm} 
\end{figure}

In the most deeply detached conditions, the electron density near the target is expected to become smaller, as further drops in temperature promote increasingly strong ion sinks when electron-ion recombination is the dominant ion sink near the target. The behaviour of the electron density profile under more deeply detached conditions can be studied by comparing the NBI heated core density ramp ($P_{SOL} \sim 1.2$ MW) to an Ohmic density ramp ($P_{SOL} \sim 0.6$ MW), where no additional external heating source is used during the discharge. The electron density profile along the separatrix during the Ohmic core density ramp \#48140 is shown in figure \ref{fig:fig10_48140_sep}.
The first difference that can be noticed is the generally lower electron density, correlated to the lower power crossing the separatrix. The density profile is initially peaked at the target, as observed in the most detached stages of the NBI heated ramps. It then slightly increases before the density front detaches from the target and moves further upstream, leaving a region of low density behind ( $\leq 1 \cdot 10^{19}$ m$^{-3}$), consistent with spectroscopic analysis \cite{Verhaegh_2023}. 
\section{Discussion}
The inferred electron density profiles can now be compared to different markers of detachment, as well as to reduced models and SOLPS simulations, to explore the main physics processes driving the evolution of the density profile in the detached region of the divertor.
\label{sec:discussion}
\subsection{Density front detachment}\label{sec:density_detach}
\begin{figure}[]\centering
\includegraphics[width= 0.45\textwidth]{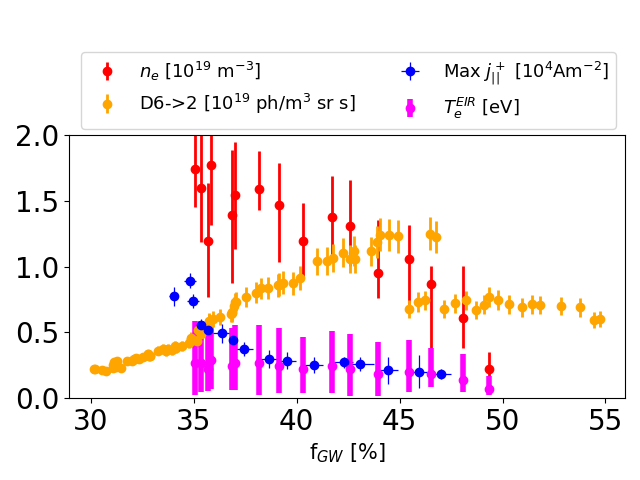}
\vspace{-0.3 cm}\caption{Comparison of the electron density and D 6$\rightarrow$2 Balmer emission (proxy for the electrion-ion recombination) at 3 cm from the target along the separatrix in Ohmic core density ramp \#48140, along with the electron temperature inferred from the two measurements and the peak ion saturation current measured by the Langmuir probes. The increase in emission while the $n_e$ has already started decreasing suggests a delayed detachment of the emisison front compared to the density front.}
\label{fig:fig11_detach_comp}
\vspace{-0.3cm} 
\end{figure}
 In the early stages of detachment ($f_{GW} \lesssim 40 \%$) the density profile is non-monotonic when expressed as a function of poloidal distance from the X-point. The maximum along the separatrix appears to be loosely correlated to the Fulcher emission front ($\sim$ detachment front) position in the beam-heated discharges. This is similar to what was observed in predictive SOLPS-ITER simulations of the SXD divertor with ion-molecule collisions turned off \cite{myatra_solps-iter_2023}, where the electron density profile peaks at the detachment front and then starts decreasing toward the target. As the divertor becomes more detached, the experimentally inferred density profiles flatten, becoming more peaked toward the target. This is more in agreement with the set of simulations that includes the ion-molecule collisions, which result in a flattening of the density profile near the detachment front and a build-up of density near the target.  An increased influence of ion-molecule collisions in later stages of the discharge could be expected by the constantly increasing neutral pressure in the divertor during the density ramp. In simulations, the density build-up is also associated with a drop in the plasma parallel velocity, which will be topic of future studies using coherence imaging flow measurements. Eventually, the density peak starts moving towards the X-point.
 
This is reminiscent of behaviours observed in JET and TCV, where the density roll-over at the target is observed with a delay with respect to the ion flux rollover \cite{lomanowski_Stark_fit_2015}\cite{VERHAEGH_TCV}, suggesting that the peak in electron density profile should be downstream of the detachment front. A significant difference is that in JET the region of peak density along the profile moves from the target to the X point shortly after the detachment onset, resulting in a decreasing density from the X point to the  target\cite{lomanowski_Stark_fit_2015}\cite{KARHUNEN_JET}. Instead, in MAST-U the density profile remains peaked at the target for most of the scanned range in upstream density, as shown in previous sections of this paper, and only detaches from the target in the most strongly detached conditions with no additional input power. 

The detachment of the density front in Ohmic conditions can be compared to the four stages of detachment defined by Verhaegh et al. in similar Ohmic density ramps \cite{Verhaegh_2023_2}. In particular, as shown in figure \ref{fig:fig11_detach_comp}, the detachment of the density front ( $\sim$ 37 $\%$ $f_{GW}$) precedes the detachment of the 6$\rightarrow$2 Balmer line emission front at $\sim$ 43 $\%$ $f_{GW}$ (as measured by the MWI diagnostic \cite{feng_development_2021}\cite{wijkamp_MWI_2023}), here used as a proxy for the electron-ion recombination (EIR) region. The peak particle flux to the target is monotonically decreasing as the core density is increased, with a steeper drop correlated to the detachment of the electron density peak. The electron temperature, inferred using the electron density from CIS and assuming that the Balmer emission is purely due to EIR, is also shown. It is obtained as a solution of
\begin{equation}
    \epsilon_{D6\rightarrow2}(n_e, T_e) = n_e^2PEC_{rec}(n_e, T_e)
    \label{eq:EIR}
\end{equation}
with the PEC effective emission coefficients taken from the ADAS database\footnote{The PEC rates in the standard ADAS database only extend down to $T_e = 0.2$ eV. The rates have been extrapolated to $T_e$ $\leq$ 0.2 eV. \cite{Verhaegh_2023_2}}.
The significant delay between the detachment of the electron density and EIR fronts is attributed to the conditions of the MAST-U divertor, with low electron densities and low electron temperatures ($\leq$ 0.5 eV). In these conditions, the amount of EIR is mostly driven by the low temperatures, as also observed in predictive simulations \cite{myatra_solps-iter_2023}, instead of by the dependency on electron density. As the divertor becomes more detached, the electron density front starts moving away from the target, which drives down the amount of the EIR, but the simultaneous decrease in electron temperature drives more EIR and associated emission. The balance of the two effects results in a delayed detachment of the EIR front compared to the electron density front. This delay might be smaller in higher density conditions, where the amount of EIR is more sensitive to its density dependence ($\propto n_e^3$)\cite{LIPSCHULTZ_recomb}.
\subsection{Comparison with analytical models}\label{sec:analytical_models}
\begin{figure}[]\centering
\begin{subfigure}[b]{0.45\textwidth}
\includegraphics[width= \textwidth]{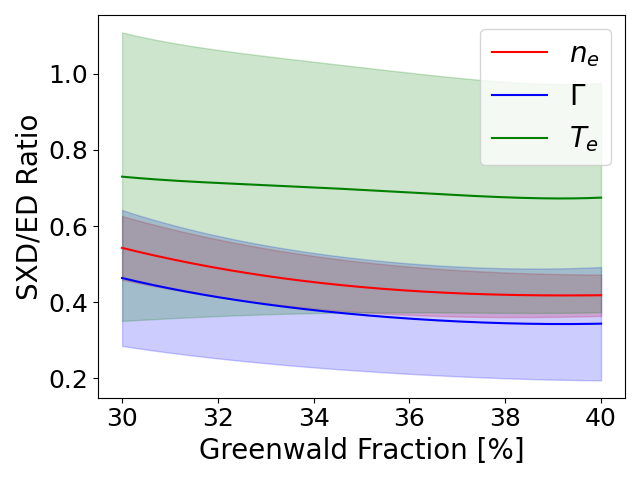}
\caption{}
\end{subfigure}
\begin{subfigure}[b]{0.45\textwidth}
\includegraphics[width= \textwidth]{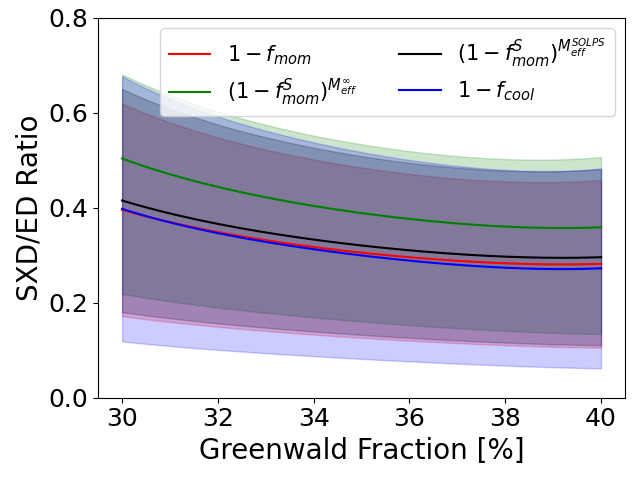}
\caption{}
\end{subfigure}
\vspace{-0.3 cm}\caption{Ratio of inferred target quantities (a) and integrated momentum and power losses (b) in the NBI heated SXD (\#46893) and ED (\#47079 ) core density ramps.}
\label{fig:fig11_f_mom_cool}
\vspace{-0.3cm} 
\end{figure}

The behavior of the density near the target can be used in conjunction with reduced analytical models to infer the  increase in total power and momentum losses with increasing total flux expansion and the volume available for plasma-neutral interactions. The modified 2-point model \cite{Stangeby_2018} can be used to relate upstream conditions, characterized by the parallel heat flux $q_{||,u}$ and total pressure $ p_{tot,u}$, to the target electron density, temperature and particle flux through power and momentum balance along a flux tube. Power and momentum losses along the flux tube (e.g. from volumetric and cross-field transport effects ) are expressed through fractional loss terms, $f_{cool}$ and $f_{mom}$ respectively.
\begin{align}
    & T_t^{2PM} = \frac{8m_i}{\gamma^2}\frac{q_{||,u}^2}{p_{tot,u}^2}\frac{\left( 1 - f_{cool}\right)^2}{\left( 1 - f_{mom}\right)^2}\left(\frac{R_u}{R_t} \right)^2  \label{eq:2PM_temp}\\
     & n_t^{2PM} = \frac{\gamma^2}{32m_i}\frac{p_{tot,u}^3}{q_{||,u}^2}\frac{\left( 1 - f_{mom}\right)^3}{\left( 1 - f_{cool}\right)^2}\left(\frac{R_u}{R_t} \right)^{-2} \label{eq:2PM_dens}\\
     & \Gamma_t^{2PM} = \frac{\gamma}{8m_i}\frac{p_{tot,u}^2}{q_{||,u}}\frac{\left( 1 - f_{mom}\right)^2}{\left( 1 - f_{cool}\right)}\left(\frac{R_u}{R_t} \right)^{-1}  \label{eq:2PM_gamma}
\end{align}
The effects of total flux expansion are explicitly accounted for by the ratio in the upstream and target major radii, $R_u$ and $R_t$, assuming that the magnetic field strength decays radially $B \propto R^{-1}$, which is well met in the equilibria considered for this work.
The lack of simple models for $f_{cool}$ and $f_{mom}$ during detachment makes a comparison of the absolute density measurements against the model difficult. However, these losses incurred before reaching the target can be estimated experimentally by putting together the electron density measured near the target by the CIS with the peak particle flux measured by the target Langmuir probes. Taking the ratio of eq (\ref{eq:2PM_dens}) in SXD and ED, assuming the same $q_{||,u}$ and $ p_{tot,u}$ in both discharges when compared for the same upstream density, and substituting it in the ratio of eq (\ref{eq:2PM_gamma}) for the two configurations gives 
\begin{align}
   &  \frac{(1-f_{cool})_{SXD}}{(1-f_{cool})_{ED}} = \frac{\Gamma_{SXD}^3}{\Gamma_{ED}^3}\frac{n_{e,SXD}^{-2}}{n_{e,ED}^{-2}}\frac{R_{SXD}}{R_{ED}} \\
    & \frac{(1-f_{mom})_{SXD}}{(1-f_{mom})_{ED}} = \frac{\Gamma_{SXD}^2}{\Gamma_{ED}^2}\frac{n_{e,SXD}^{-1}}{n_{e,ED}^{-1}} \label{eq:2PM_mom_loss}
\end{align}
These equations express the ratio in power and momentum reaching the target in the SXD compared to the ED.
 Similar equations can be found using an extension of the 2-point model that explicitly accounts for possible momentum losses due to total flux expansion in the presence of significant plasma flows \cite{Carpita_2024}. An increase in these losses has been proposed as a possible explanation for the weaker-than-expected particle flux decrease with increasing strike point radius on TCV \cite{Carpita_2024}. In this case, the momentum loss term can be divided between a term accounting for volumetric and cross-field momentum losses $(1-f^S_{mom})$ and a term accounting for geometric losses, which is a function of the effective Mach number $M_{eff}\cite{Carpita_2024}$, thus
\begin{align}
    & (1-f_{mom}) = (1-f^S_{mom})\left(\frac{R_{u}}{R_{t}}\right)^{\frac{M_{eff}^2}{1 + M_{eff}^2}}  
    \end{align}
Assuming the value of $M_{eff}$ does not vary significantly between the ED and SXD configurations, the ratio of the volumetric and cross-field momentum losses can be expressed explicitly as 
\begin{align}
    \frac{(1-f^S_{mom})_{SXD}}{(1-f^S_{mom})_{ED}} = \frac{\Gamma_{SXD}^2}{\Gamma_{ED}^2}\frac{n_{e,SXD}^{-1}}{n_{e,ED}^{-1}}\left(\frac{R_{SXD}}{R_{ED}} \right)^{\frac{M_{eff}^2}{1 + M_{eff}^2}}  \label{eq:mom_frac_Meff}
\end{align}
While experimental values of $M_{eff}$ are not available, 3 limiting cases can be considered:
\begin{itemize}
    \item A case where the effect of flows are not significant ($M_{eff} = 0$), leading to no additional momentum losses and the original 2PM equation (\ref{eq:2PM_mom_loss})
    \item A case using a characteristic $M_{eff}$ value computed from interpretative SOLPS simulations of the discharges (explored more in section \ref{sec:SOLPS})  ($M_{eff}^{SOLPS} \sim 0.5$) 
    \item A case which maximises the effect of geometric losses and minimizes the volumetric effects ($M_{eff}^{\infty} \rightarrow \infty$)
\end{itemize}
The inferred target electron density and particle flux ratios between the SXD and ED during the NBI heated core density ramps are shown in figure  \ref{fig:fig11_f_mom_cool}a as a function of core Greenwald fraction, along with the $T_e$ ratio that can be inferred from the two measurements, assuming sonic flows at the target as in the traditional 2 point model:
\begin{align}
    \frac{T_e^{SXD}}{T_e^{ED}} = \frac{\Gamma_{SXD}^2}{\Gamma_{ED}^2}\frac{n_{e,SXD}^{-2}}{n_{e,ED}^{-2}} \label{eq:CIS_LP_Te_inf}
\end{align}
The target electron density and particle flux are lower in the SXD by about 50 \% and 60 \% compared to their values in ED respectively. Therefore, based on equations \ref{eq:mom_frac_Meff} and \ref{eq:CIS_LP_Te_inf}, the inferred power $(1-f_{cool})$ and momentum $(1-f_{mom})$ reaching the target are $\sim$ 60 - 70 \% smaller in SXD than the ones in ED, as shown in figure \ref{fig:fig11_f_mom_cool}b, evidence of the significant improvement in performance with increasing poloidal leg-length and total flux expansion.

The particle flux, electron density, power loss and momentum loss ratios all show a slightly decreasing trend towards higher upstream densities. This suggests that the increased benefits from long-legged configurations are stronger in higher divertor neutral pressure conditions, probably due to the increased role of volumetric losses. 

The strong increase in momentum losses in MAST-U is much larger than what would be expected from geometric effects alone. Using  $M_{eff}^{SOLPS}$ leads to $\sim$ 95 \% of the increase in momentum losses being attributed to volumetric and cross-field effects, while even in the limit case $M_{eff}^{\infty}$ this value is still of $\sim$ 79 \%. This suggests that in MAST-U the momentum losses are dominated by volumetric and cross-field effects, compared to TCV in which the geometric term may play a more dominant role. A possible explanation of this difference is the stronger baffling on MAST-U, increasing the strength of the volumetric effects and reducing the neutral leakage to the midplane, and thus the flow toward the divertor of the neutrals ionized in the main chamber.
\begin{figure}[]\centering
\begin{subfigure}[b]{0.45\textwidth}
\includegraphics[width= \textwidth]{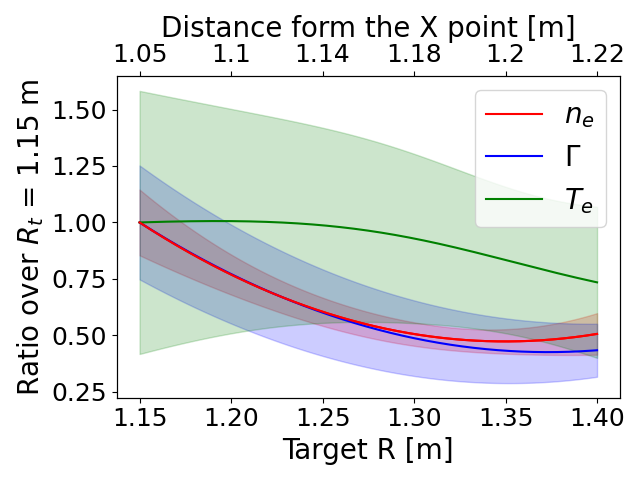}
\caption{}
\end{subfigure}
\begin{subfigure}[b]{0.45\textwidth}
\includegraphics[width= \textwidth]{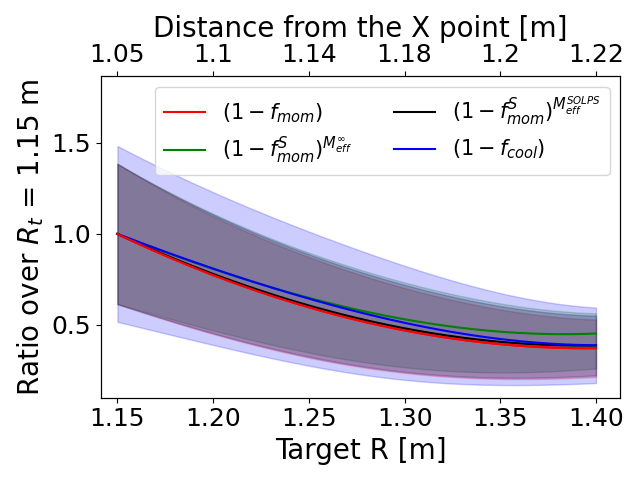}
\caption{}
\end{subfigure}
\vspace{-0.3 cm}\caption{Ratio of inferred target quantities (a) and integrated momentum and power losses (b) as the leg is swept from ED to a larger target radius in NBI heated strike point scan \#46895.}
\label{fig:fig12_f_mom_cool_scan}
\vspace{-0.3cm} 
\end{figure}

A similar type of analysis can be applied to the poloidal leg length scan with constant upstream conditions. In this case, the inferred values are normalized to the initial value of the scan, with the strike point major radius at 1.15 m. As the leg is swept out, both the target electron density and the particle flux are observed to decrease by $\sim $50 \%, as shown in figure \ref{fig:fig12_f_mom_cool_scan}a.
The momentum and power reaching the target are shown in figure \ref{fig:fig12_f_mom_cool_scan}b and they are inferred to strongly decrease from their initial values as the leg is swept out, leading to a $\sim$ 60 \% reduction in both power and momentum.
In this case, volumetric and cross field momentum losses are inferred to dominate over momentum losses related to flows over the entire scan, with average values of 97 \% and 90 \% of the momentum losses in the M$_{eff}^{SOLPS}$ and M$_{eff}^{\infty}$  cases respectively.
\subsection{Comparison with SOLPS simulations}
\label{sec:SOLPS}
\begin{figure}[]\centering
\begin{subfigure}[b]{0.48\textwidth}
\includegraphics[width= \textwidth]{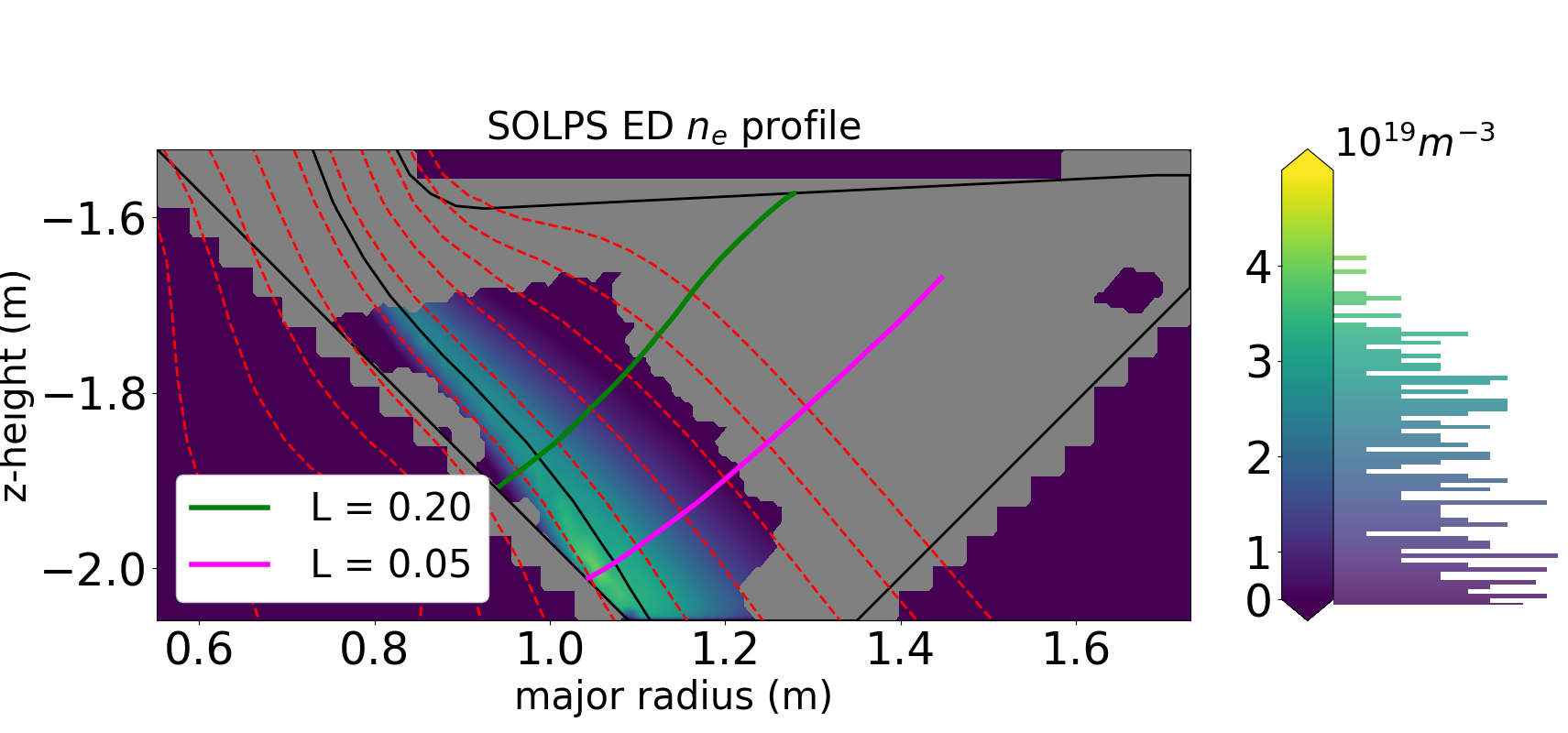}
\caption{}
\end{subfigure}
\begin{subfigure}[b]{0.48\textwidth}
\includegraphics[width=\textwidth]{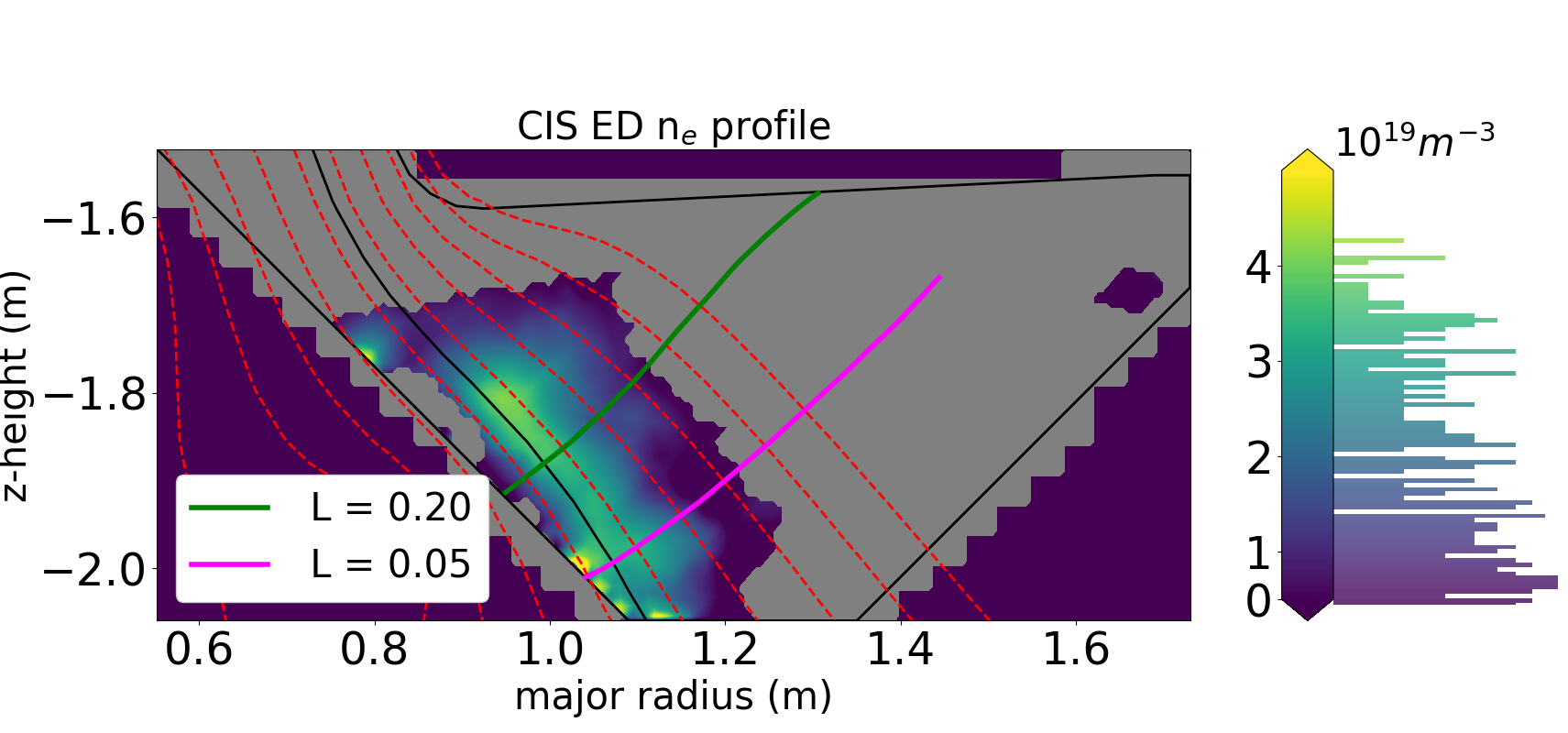}
\caption{}
\end{subfigure}
\begin{subfigure}[b]{0.35\textwidth}
\includegraphics[width= \textwidth]{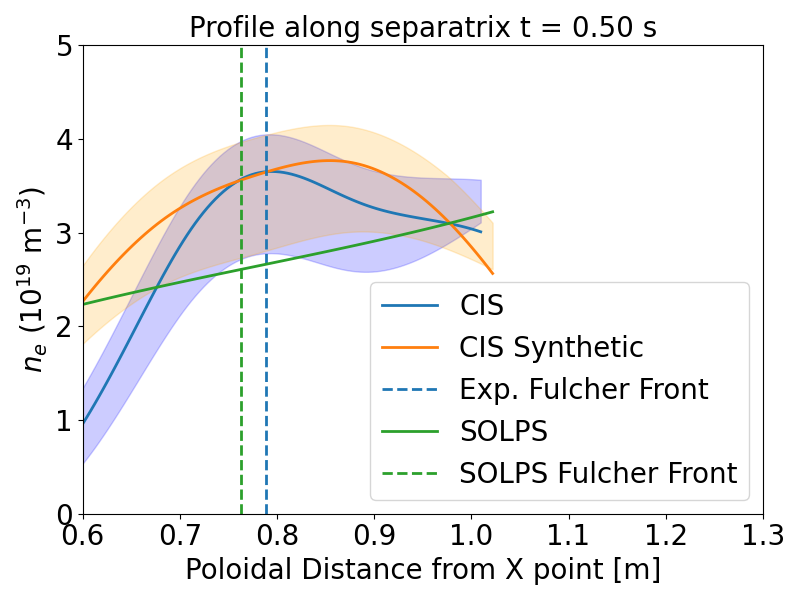}
\caption{}
\end{subfigure}
\vspace{-0.3 cm}\caption{ (a) $n_e$ profile from an interpretative SOLPS simulation of ED NBI-heated discharge \#47079. (b) Experimental profile matching the same upstream density. A histogram of the plotted values is shown along the colorbars. (c) Comparison of experimental, simulated, and synthetic electron density profiles along the separatrix. The experimental and modeled fulcher emission front are shown as vertical lines.}
\label{fig:fig13_SOLPS_ED}
\vspace{-0.3cm} 
\end{figure}

The inferred poloidal electron density profiles can be compared against the profiles from interpretative SOLPS simulations of these discharges\cite{verhaegh2024improved}\footnote{The SXD simulation is based on experimental discharge \#46860, of which \#46893 is a repeat discharge.} to validate the capabilities of the code in the strongly baffled long-legged conditions available on MAST-U. To account for possible artifacts and systematic errors that can appear during the tomographic inversion to obtain the 2D profiles, the measurements are also compared to "synthetic" 2D profiles (with their respective uncertainties) obtained by using the SOLPS profiles to generate synthetic CIS data, which is then tomographically inverted as previously done to characterize the performance of the diagnostic \cite{lonigro_CIS}. 
The experimental times corresponding to the simulations are chosen by matching the midplane electron density, measured by Thomson scattering at the separatrix inferred by the EFIT magnetic reconstruction, to the midplane electron density in SOLPS, and roughly corresponds to the start of the density ramps. The SOLPS simulations are in a comparable depth of detachment as experiment, as can be noticed by the similar position of the modeled and experimental fulcher emission front location(proxy for the 3 eV front \cite{Verhaegh_2023}), shown in figure \ref{fig:fig13_SOLPS_ED}c and \ref{fig:fig16_SOLPS_SXD_sep}.  

\begin{figure}[]\centering
\begin{subfigure}[b]{0.45\textwidth}
\includegraphics[width= \textwidth]{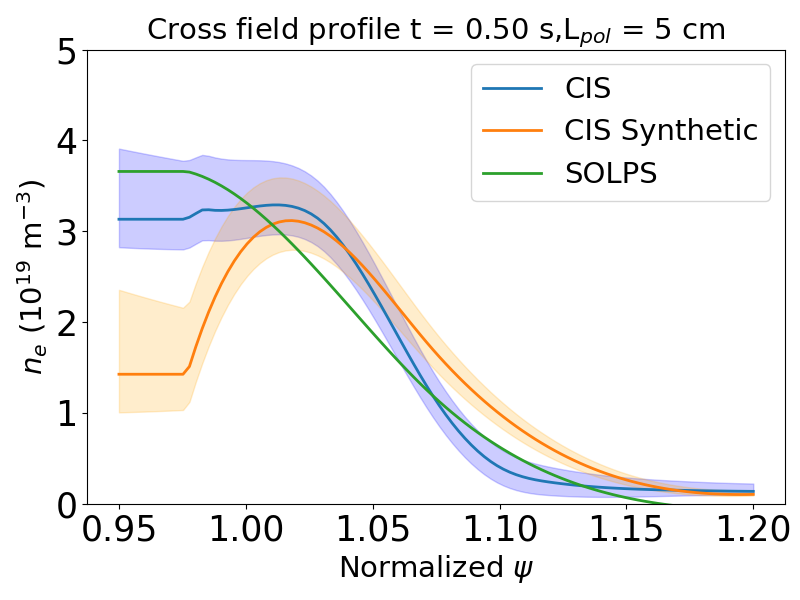}
\caption{}
\end{subfigure}
\begin{subfigure}[b]{0.45\textwidth}
\includegraphics[width= \textwidth]{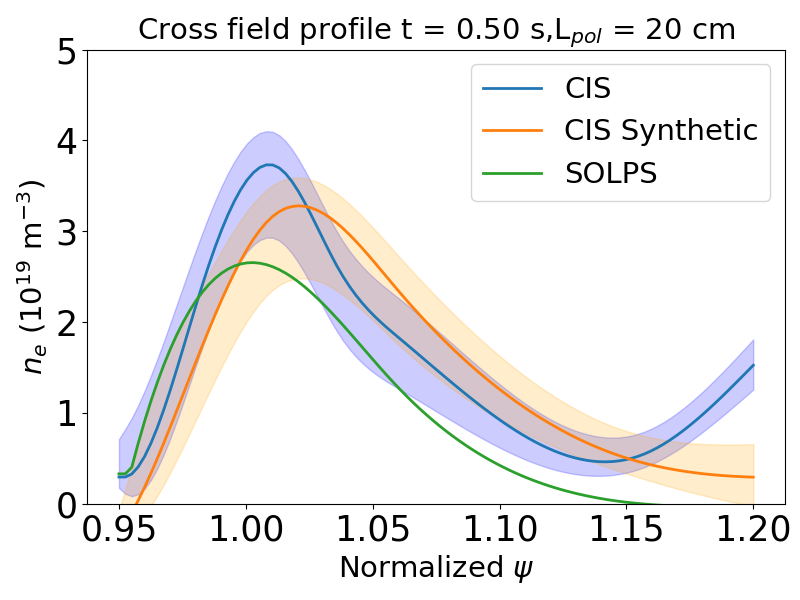}
\caption{}
\end{subfigure}
\vspace{-0.3 cm}\caption{Comparison of experimental, simulated, and synthetic electron density profiles across the separatrix at 5 cm (a) and 20 cm (b) from the target in ED NBI-heated discharge \#47079 .}
\label{fig:fig14_SOLPS_ED_cross}
\vspace{-0.3cm} 
\end{figure}

The 2D $n_e$ profile obtained from the interpretative simulation, including drifts, of ED discharge \#47079 with a simulated power crossing the separatrix of P$_{SOL}^{SOLPS} = 1$ MW, is shown in figure \ref{fig:fig13_SOLPS_ED}. Also shown is the comparison between the experimental and simulated electron density profiles along the separatrix. 
Analogously, the cross-field profiles are compared at 5 and 20 cm from the target in figure \ref{fig:fig14_SOLPS_ED_cross}.
Overall, there is good agreement with the interpretative simulation in both the magnitude and shape of the profiles. While the experimental profile is inferred to be slightly decreasing downstream of the density maximum near the detachment front, the profile is still in agreement within uncertainty with the simulated profile which is peaked at the target. While the agreement between experimental and simulated profiles is reasonable, the cross-field synthetic SOLPS profile shows an underestimation of the density in the PFR near the target. That suggests a possible limitation of the camera view when trying to reconstruct the behavior in the narrow gap between the separatrix and the tile close to the target in this magnetic geometry and thus the inference results may be unreliable in that region. 

\begin{figure}[]\centering
\begin{subfigure}[b]{0.45\textwidth}
\includegraphics[width= \textwidth]{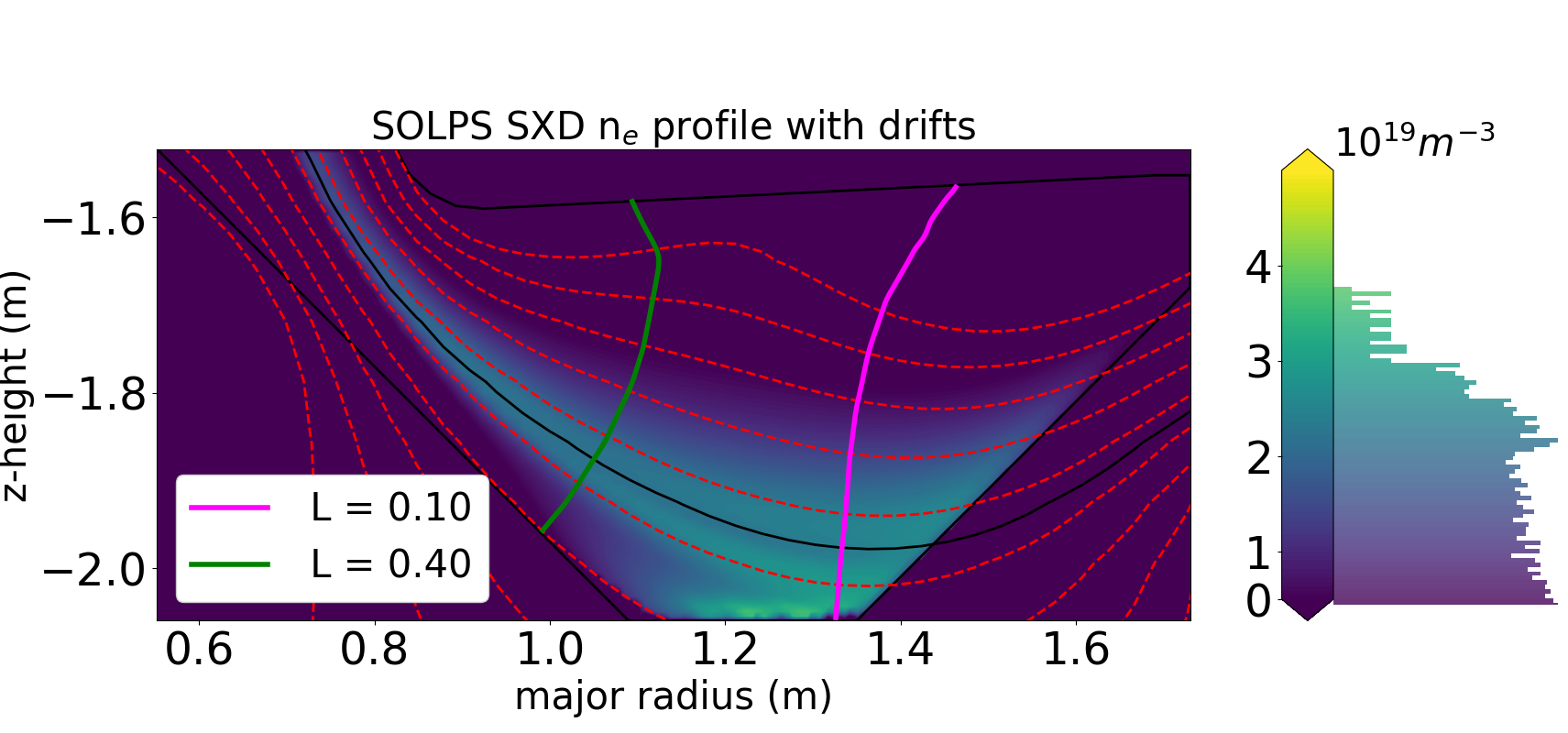}
\caption{}
\end{subfigure}
\begin{subfigure}[b]{0.45\textwidth}
\includegraphics[width= \textwidth]{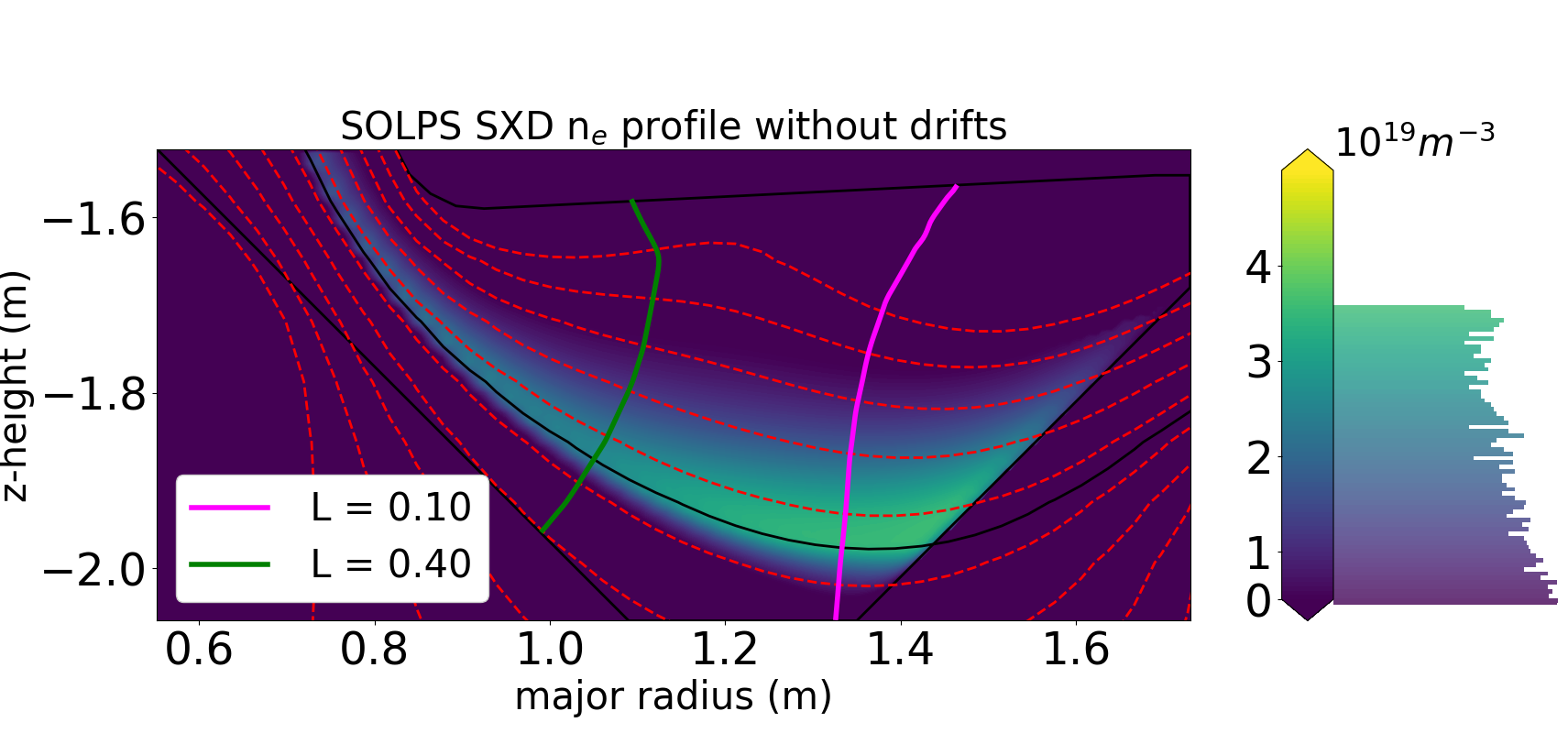}
\caption{}
\end{subfigure}
\begin{subfigure}[b]{0.45\textwidth}
\includegraphics[width=\textwidth]{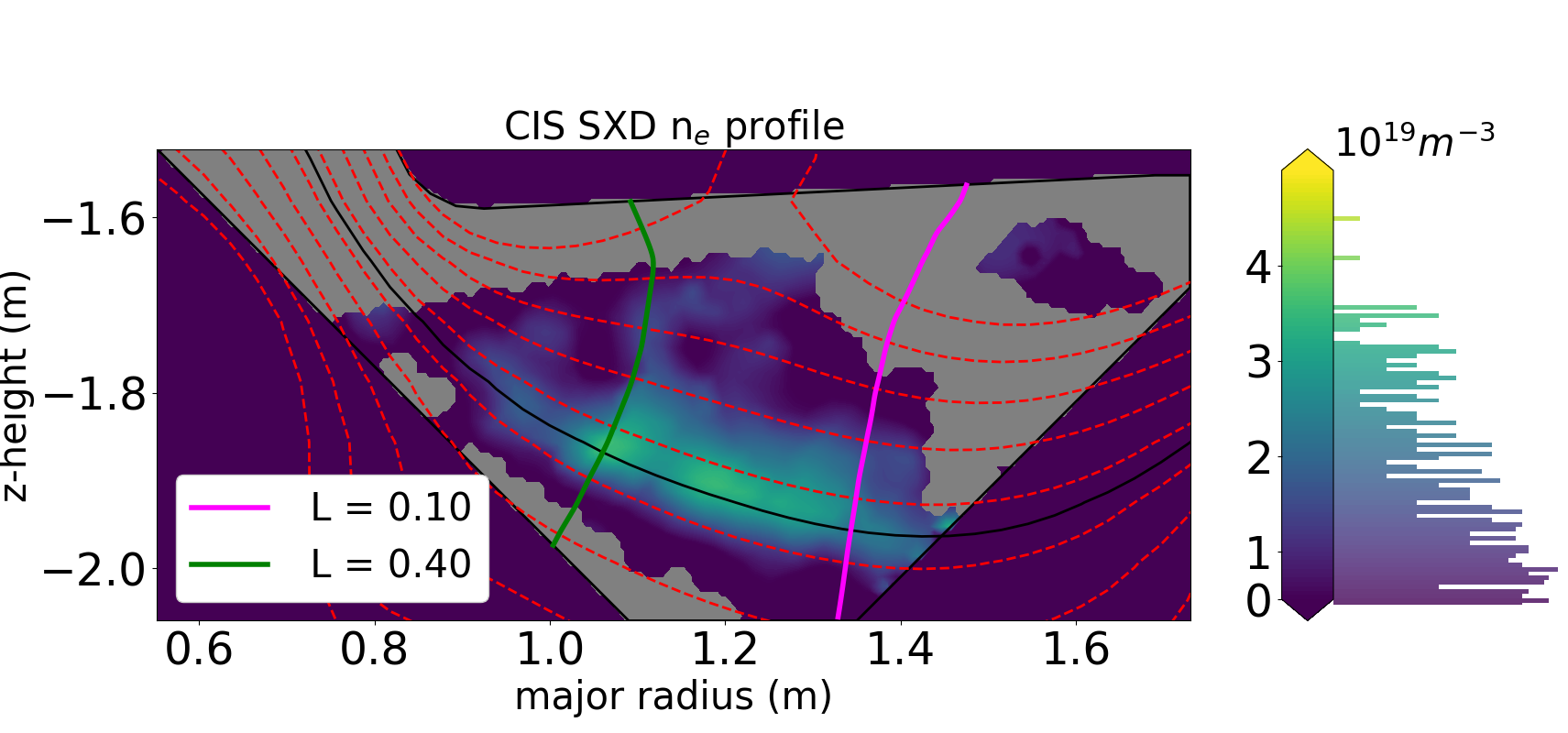}
\vspace{-0.3 cm}
\caption{}
\end{subfigure}
\caption{2D $n_e$ profiles from interpretative simulations of SXD NBI-heated discharge \#46860 with drifts (a) enabled and (b) disabled. (c) Experimentally inferred profile of NBI-heated SXD discharge repeat \#46893. A histogram of the plotted values is shown along the colorbars.}
\label{fig:fig15_SOLPS_SXD_2D}
\vspace{-0.3cm} 
\end{figure}

\begin{figure}[]\centering
\includegraphics[width= 0.45\textwidth]{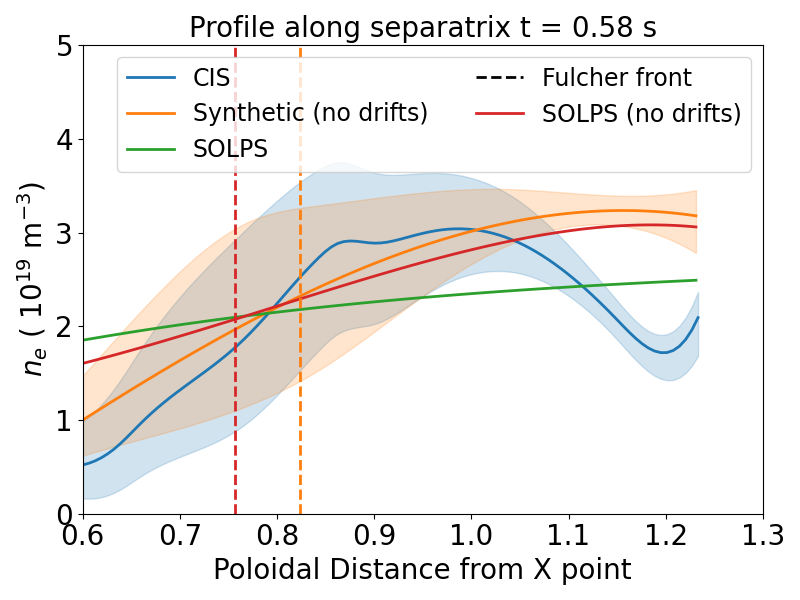}
\caption{Comparison of experimental (\#46893), simulated, and synthetic electron density profiles along the separatrix for SXD NBI-heated discharge \#46860. The experimental and modelled fulcher emission front are shown as vertical lines.}
\label{fig:fig16_SOLPS_SXD_sep}
\vspace{-0.3cm} 
\end{figure}

\begin{figure}[]\centering
\begin{subfigure}[b]{0.45\textwidth}
\includegraphics[width= \textwidth]{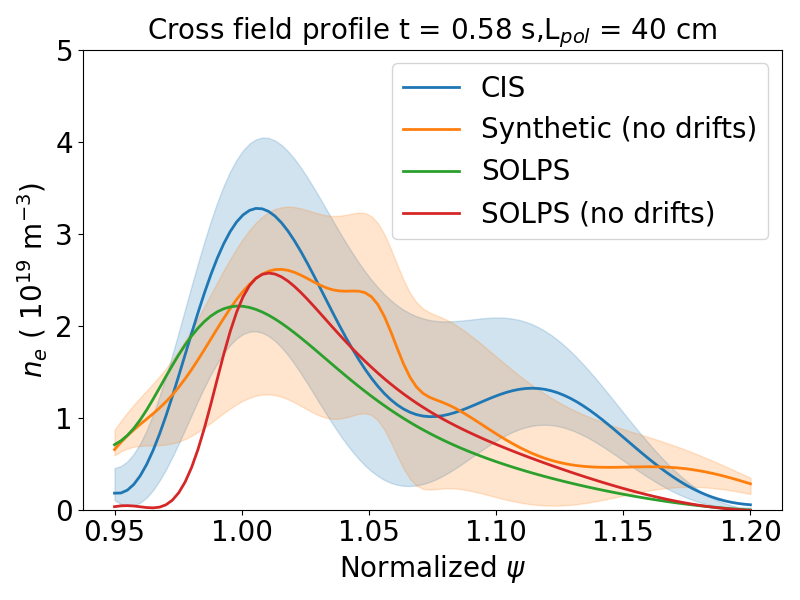}
\caption{}
\end{subfigure}
\begin{subfigure}[b]{0.45\textwidth}
\includegraphics[width= \textwidth]{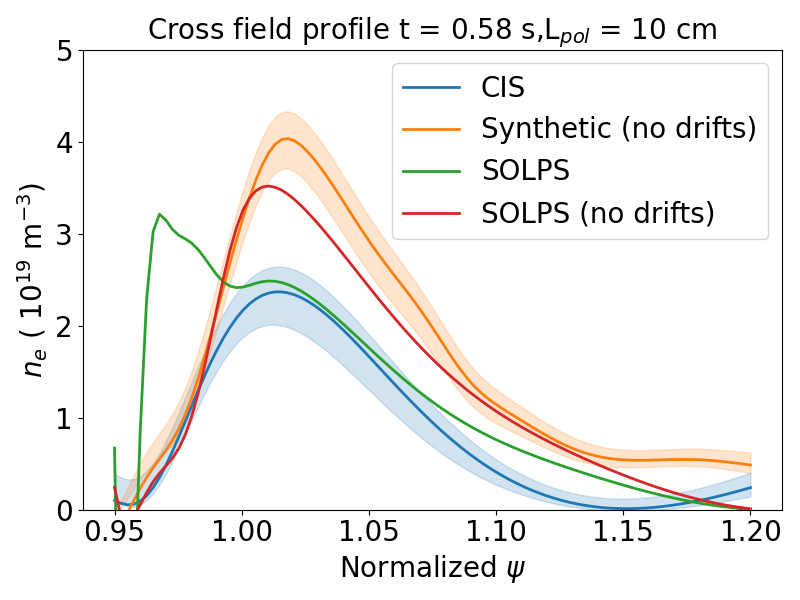}
\caption{}
\end{subfigure}
\vspace{-0.3 cm}
\caption{Comparison of experimental (\#46893), simulated (without drifts,\#46860), and synthetic electron density profiles across the separatrix at 40 cm (a) and 10 cm (b) from the target in an SXD NBI-heated discharge.}
\label{fig:fig17_SOLPS_SXD_cross_ield}
\vspace{-0.3cm} 
\end{figure}

\begin{figure}[]\centering
\begin{subfigure}[b]{0.45\textwidth}
\includegraphics[width= \textwidth]{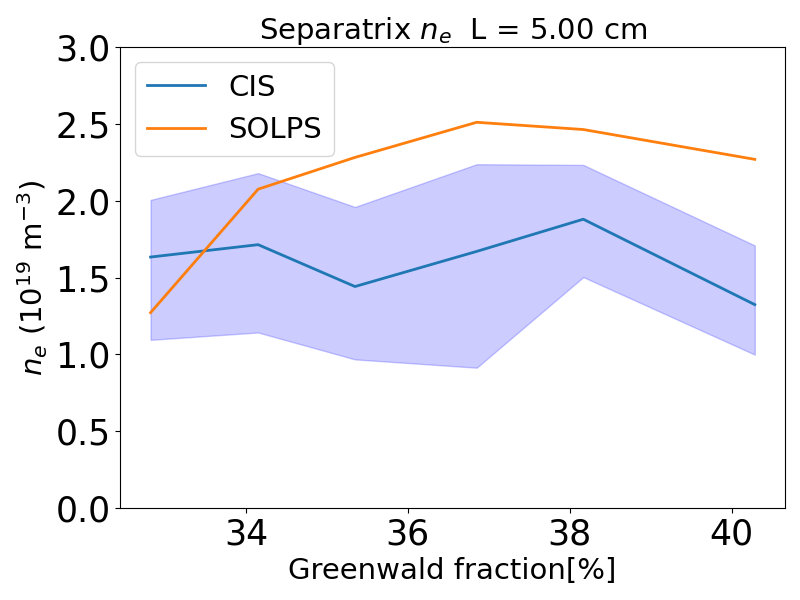}
\caption{}
\end{subfigure}
\begin{subfigure}[b]{0.45\textwidth}
\includegraphics[width= \textwidth]{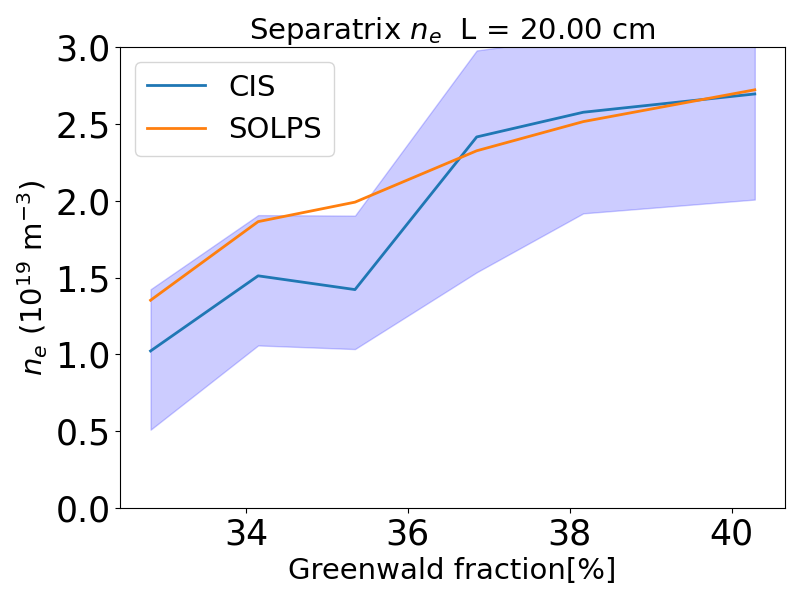}
\caption{}
\end{subfigure}
\vspace{-0.3 cm}
\caption{Comparison of inferred and simulated electron density along the separatrix at 5 (a) and 20 (b) cm from the target as a function of upstream density in SXD Ohmic discharge \# 48140.}
\label{fig18:SOLPS_Ohmic_ramp}
\vspace{-0.3cm} 
\end{figure}

The electron density profile is less in agreement in the SXD case, compared to both a simulation with and without drifts enabled, as shown in figure \ref{fig:fig15_SOLPS_SXD_2D}. Given the density peak near the bottom of the divertor, not observed in experiment, the simulation without drifts has been used to generate the synthetic CIS profiles, as overall the spatial profile has been deemed in better agreement. \\
The peak density in the experimental profile is comparable to the peak density in the simulation without drifts, while being larger than in the simulation with drifts. 
The spatial profile along the separatrix is also different, with the experimental inference showing a peak in the middle of the chamber and then decreasing toward the target, while the simulations showing either a flat profile (with drifts) or a profile peaked at the target (without drifts), as shown in figure \ref{fig:fig16_SOLPS_SXD_sep}. Better agreement is obtained if the simulated profiles are compared to later times during the core density ramps, as the experimental density profiles flattens in the divertor chamber, although the inferred non-monotonic profile is stll not observed in simulations of more attached conditions.\\
Another difference is in the extent of the cross-field profiles, where the simulation with drifts predicts a broader profile with significant spreading in the PFR, leading to a build-up of density near the bottom of the divertor. While a shift of the density profile toward the PFR is observed in experiment, no peak in density near the bottom tile is inferred by the CIS. Furthermore, the existence of such a significant density peak in the PFR would also lead a peaked Balmer emission profile near the bottom of the divertor, also not observed. The cross-field profiles are compared in figure \ref{fig:fig17_SOLPS_SXD_cross_ield} at a parallel distance from the target ($L_{pol}$) of 10 cm and 40 cm. At 40 cm from the target, the profile is in reasonable agreement with both simulations, in particular regarding the sharp drop in the PFR and the peak density. Closer to the target ($L_{pol}$ = 10 cm), the peak density is in better agreement with the simulation with drifts, although the peaking of the density near the bottom tile is not observed. The profile width in the far SOL is comparable between simulations and experiment. 
This suggests that the effect of drifts in these simulations of deeply-detached SXD geometries may introduce some discrepancies with experimental data. Future comparison in attached conditions could indicate if this is related specifically to the longer-legged geometry or to the temperature range achieved in these detached conditions, in which MAR particle sinks may be underestimated \cite{Verhaegh_CX}.\\
More deeply detached conditions can also be compared to experiment using the Ohmic discharge (\#48140) as a reference. Here, inferences of the 2D electron density profiles are complicated by the lower electron densities, increasing the uncertainty of the Stark-broadening-based measurements, and thus making a detailed comparison of the profiles from SOLPS and CIS along and across the separatrix less reliable. Furthermore, numerical instabilities are observed in the SOLPS simulations after the electron density front detaches from the target. Therefore, the comparison of experimental measurements of density with SOLPS predictions has been limited to the values on the separatrix and close to the target, where the densities are highest and the inference more reliable, to compare the early stages of the density front detachment from the target observed in Ohmic conditions with the simulations. The inferred electron densities along the separatrix at 5 and 20 cm from the target are compared with increasing upstream electron density to a series of SOLPS simulations without drifts, with P$_{SOL}$ = 0.6 MW, and increasing fuelling rate in figure \ref{fig18:SOLPS_Ohmic_ramp}.
The inferred divertor electron densities are in reasonable agreement throughout the density scan up to the density front detachment, albeit the density very close to the target is slightly underestimated, as observed in the NBI-heated case. This may suggest improved agreement in lower electron temperature conditions, dominated by EIR\\

\subsection{Neutral drag and electron density build-up near the target}
Trying to understand possible reasons that could explain the discrepancies between experiment and SOLPS simulations in the NBI heated SXD case (figure \ref{fig:fig16_SOLPS_SXD_sep}) is instructive to determine the dominant physical processes at play in setting the electron density profile downstream of the detachment front. In turn, the density profile can then affect the amount of power and particle losses in the detached region of the plasma and affect the resulting particle and heat fluxes at the target.

The evolution of the electron density profile downstream of the detachment front during the core density ramp is in qualitative agreement with simple modeling based on the competition of neutral drag, which increases the electron density, and recombination ion sinks, which decrease it. 

Considering a 1D model along a flux tube and neglecting cross-field transport, the density at each point will be given by the solution of the continuity equation. Focusing on the region downstream of the detachment front, where negligible ionization is assumed, the density at each point (T) will be a function of the density at the detachment front (DF), the integrated volumetric recombination between the front and the point and the flow velocities at the detachment front and the point T. The effect of non-constant cross-field area ($A$) can also be accounted for, such as in the presence of total flux expansion along the flux surface:
\begin{align}
    \frac{\dd}{\dd x}\left[n_e(x)v(x)A(x)\right] = -S_{rec}(x)A(x)
\end{align}
\begin{align}
    n_e^{DF}v^{DF}A^{DF} -  n_e^{T}v^{T}A^{T}= -\int_{T}^{DF}S_{rec}(x)A(x) \dd x \label{eq:continuity}
\end{align}
Given the lack of ionization downstream of the detachment front and the presence of strong volumetric particle sinks in the MAST-U divertor \cite{verhaegh2024nbi}, which would tend to lower the electron density, another mechanism must be present to increase it and explain the peaked density at the target in experiment and SOLPS simulations. A possible explanation is that a strong drop in parallel velocity must be present, which has also been observed in predictive SOLPS simulations of the Super-X divertor and has been attributed to ion-molecule collisions \cite{myatra_solps-iter_2023}, consistent with spectroscopic analysis of D2 molecules rotational temperature \cite{osborne2024_NBI}.

The non-monotonic behaviour of the density spatial profile along the separatrix at the start of the SXD discharge (section \ref{sec:evol_during_detach}) can then be understood as MAR recombination processes dominating over the neutral drag near the target. As the core density ramp progresses, the neutral pressure in the divertor rises, increasing the drag. Furthermore, the increase in neutral pressure is also consistent with a reduction in the electron temperature, which can also reduce the efficiency of the MAR particle sinks \cite{verhaegh_molecules_2_2021}.  This is consistent with a flattening of the electron density profile or even a peaking at the target. Lowering $T_e$ also promotes electrion-ion recombination in later stages of the discharge \cite{verhaegh2024nbi}. Eventually, the electron-ion recombination losses become strong enough to dominate over the effect of the neutral drag and cause a reduction of the electron density near the target and the movement of the electron density peak away from the target, as observed experimentally in Ohmic conditions (figure \ref{fig:fig10_48140_sep}).

To try and show this behaviour qualitatively, the continuity equation can be solved numerically between the detachment front and the target using effective ion recombination rates for MAR and EIR and taking information on the neutral drag from the SOLPS simulations. Details on this simplified model are given in \ref{sec:appendix_1D_model}.

\begin{figure}[]\centering
\begin{subfigure}[b]{0.45\textwidth}
\includegraphics[width= \textwidth]{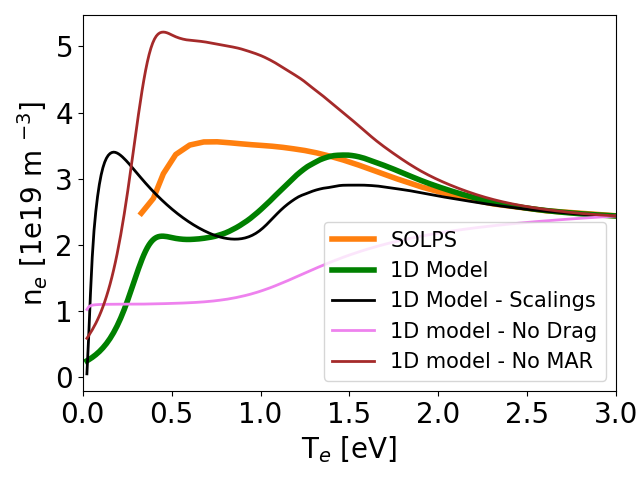}
\caption{}
\end{subfigure}
\begin{subfigure}[b]{0.45\textwidth}
\includegraphics[width= \textwidth]{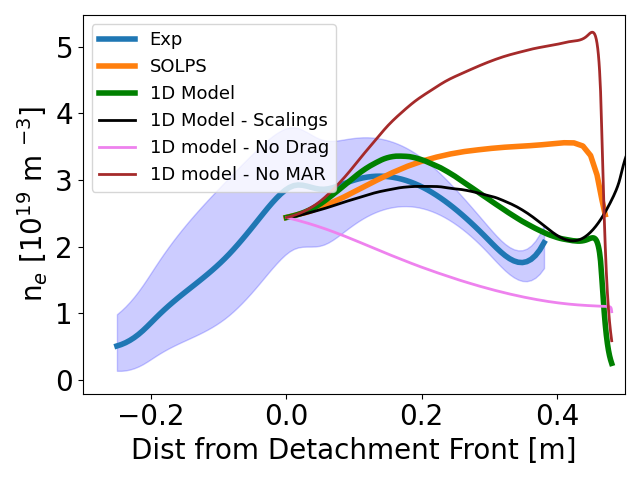}
\caption{}
\end{subfigure}
\vspace{-0.3 cm}
\caption{ Comparison of SOLPS $n_e$ profile with the results of the 1D model taking the information from SOLPS and the 1D model using the simplified scaling laws. Plotted against $T_e$ (left) and poloidal distance downstream of the detachment front (right). }
\label{fig19:1D_model}
\vspace{-0.3cm} 
\end{figure}

The solution of the equation leads to the non-monotonic behaviour shown in figure \ref{fig19:1D_model}a, with the density increasing downstream of the detachment front down to 1.5 eV and then decreasing at lower temperatures. The second peak at 0.2-0.5 eV is in an intermediate temperature range, where the MAR particle losses become less significant but the temperature is still too high for significant EIR losses at these low $n_e$ values. The profile can also be compared to the solution obtained assuming no drag (i.e. $v^{DF} = v^{T}$), where the density monotonically decreases downstream of the front, and to the solution neglecting MAR losses, which reaches significantly higher peak densities, highlighting how both of these processes are necessary to obtain peak densities along the profiles comparable to the SOLPS results.

Using the same information on the temperature gradient used to solve the equation, $\frac{\dd T}{\dd x}(T_e)$, the density curve can be expressed as a function of poloidal distance downstream of the detachment front, and thus directly compared to the experimentally inferred values, shown in figure \ref{fig19:1D_model}b.  
Although the agreement with experiment is not perfect, the comparison suggest that a model based on the competition of neutral drag and volumetric particle sinks can produce non-monotonic electron density behaviour comparable in magnitude to what is observed in experiment. If the divertor densities are higher, the volumetric particle sinks will also increase and it is reasonable to expect that this non-monotonic density behaviour would become stronger and the density would not be able to rise again near the target at intermediate temperatures ($T_e \leq 0.7$ eV) as EIR sinks would already be strong at those temperatures. 

This behaviour can be qualitatively tested by obtaining scalings for the required data to solve the continuity equation to higher densities, as detailed in \ref{sec:appendix_1D_model}. The curves obtained by solving the equation using these interpolated scalings instead of the values taken directly from SOLPS are compared to the previous curves in figure \ref{fig19:1D_model}. It can be noticed how the second density peak near the target is over-estimated, as the scaling does not fully capture the flattening of the parallel velocity very close to the target predicted by SOLPS. 

\begin{figure}[]\centering
\includegraphics[width= 0.45\textwidth]{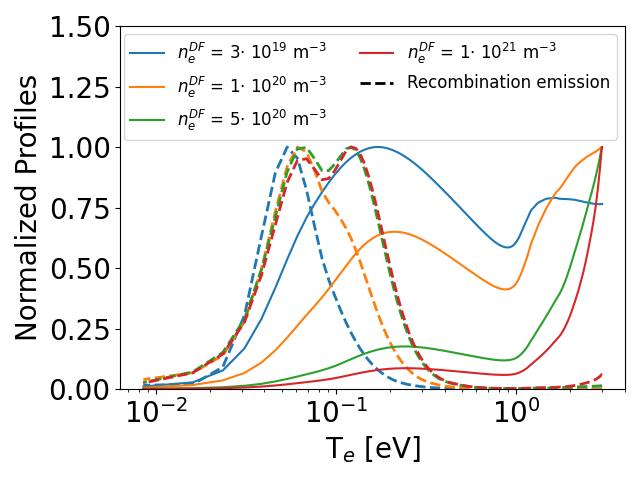}
\vspace{-0.3 cm}
\caption{ Comparison of $n_e$ and $D_\alpha$ EIR emissivity profiles normalized to their respective maxima, obtained by solving the 1D model extrapolation at higher density. The second peak at the target vanishes at higher densities due to the stronger sinks.}
\label{fig20:1D_model_extrapolation}
\vspace{-0.3cm} 
\end{figure}

Using these scalings and scanning the starting electron density at the detachment front leads to the results in figure \ref{fig20:1D_model_extrapolation}.
At low electron densities at the detachment front, the $n_e$ profile has a second peak at the target, caused by the low amount of recombination until very low temperatures are reached ($\leq$ 0.3 eV). As the detachment front densities increases, so do the ion sinks and this behaviour is not observed. The peak density along the profile is now at the first local maxima, before MAR ions sinks lead to the decrease in density. The peak density along the profile keeps moving to higher electron temperatures as the electron density increases and suggests that the density front would be easier to detach from the target in higher density machines compared to MAST-U. 

The $n_e$ and $T_e$ information can also be used to model the EIR emission for the $D_\alpha$ Balmer line using eq (\ref{eq:EIR})\footnote{This does not include possible effects of plasma-molecular interactions on the Balmer emission}. The $n_e$ and $\epsilon$ profiles normalized to their respective maxima are compared in figure \ref{fig20:1D_model_extrapolation}. At low electron densities, the EIR emission is peaked at the target and starts decreasing (i.e. detaching from the target) at lower temperatures then the density front, as observed experimentally in MAST-U. At higher detachment front densities, the EIR emission presents two peaks, one corresponding to the density peak and one at low temperatures, which would be at the target if electron temperatures $\leq$ 0.5 eV can be achieved . The extrapolated behaviours at higher densities are in qualitative agreement with the behaviour reported on JET:  a density peaking at the target at higher upstream densities (and lower temperatures) than the ion flux rollover, and a continuously decreasing target density after the maximum is reached. Furthermore, two peaks in recombination emission are observed in deep detachment, the first close to the maximum density (near the X-point) and the second one at the target driven by low electron temperatures \cite{KARHUNEN_JET}.
\section{Conclusions}
The behavior of the 2D electron density profile in the MAST-U divertor has been characterized in L-mode discharges with varying strike point radius, upstream density, and input power, highlighting the potential of CIS diagnostics to characterize the behavior of the divertor plasma parameters.

As the divertor poloidal leg length and total flux expansion are increased in strongly baffled conditions, the power and momentum losses along the separatrix are inferred to strongly increase, resulting in decreasing target electron density and particle flux and highlighting the increased exhaust performance of the Super-X divertor.  This contributes to previous studies on the benefits of long-legged and totally flux expanded divertor configurations\cite{verhaegh2024improved}\cite{moulton_super-x_2024} which might be required to operate higher power machines, such as reactors, safely.

The electron density profile downstream of the detachment front was studied as the divertor detachment state evolves and the ionization region moves away from the target. The electron density profile is inferred to peak downstream of the ionization front and exhibits a complex behaviour. It is  characterized by an initially non-monotonic profile, the building up of a density peak near the target, which then moves away from the target and towards the X-point. This behaviour is in qualitative agreement with its simplified modeling as a balance between different ion recombination processes and neutral drag acting on the plasma flow, suggesting these are the main processes determining the electron density profile downstream of the detachment front in the MAST-U conditions. Extrapolating this model to higher electron density conditions than those available on MAST-U shows that the target density peak at low temperatures vanishes, as the effect of electron-ion recombination sinks become stronger at higher electron temperatures, and that the density peak is expected to detach from the target at higher temperatures.  

Comparison of the electron density profiles inferred from the coherence imaging measurements with interpretative SOLPS-ITER simulations shows generally reasonable agreement in the electron density profiles along and across the separatrix. Some discrepancies are observed in moderately detached conditions, where the experimentally inferred non-monotonic profile along the separatrix in the early stages of the Super-X density ramp is not well reproduced by SOLPS. This is attributed to an overestimation of the effect of drag or underestimation of the particle sinks near the target, thus predicting a density profile peaked at the target. Better agreement is found in more deeply detached conditions, in which the detachment of the electron density peak from the target appears to be tracked well by SOLPS. The predicted shift of the electron density profile towards the PFR also appears stronger in SOLPS simulations compared to experiment. Further comparisons of experimental data and modeling are required to determine the cause of these possible discrepancies and how they vary with divertor geometry and depth of detachment. 

\section{Acknowledgements}
This work has been carried out within the framework of the EUROfusion Consortium, partially funded by the European Union via the Euratom Research and Training Programme (Grant Agreement No 101052200 — EUROfusion), and from EPSRC Grants EP/W006839/1 and EP/S022430/1. The Swiss contribution to this work has been funded by the Swiss State Secretariat for Education, Research and Innovation (SERI). Views and opinions expressed are however those of the author(s) only and do not necessarily reflect those of the European Union, the European Commission or SERI. Neither the European Union nor the European Commission nor SERI can be held responsible for them.

\appendix
\section{A simplified model balancing neutral drag and ion sinks} \label{sec:appendix_1D_model}

The continuity equation (eq \ref{eq:continuity}) can be expressed as a function of electron temperature through a change of variables, and assuming $A\propto R$:
\begin{align}
    n_e^{DF}v^{DF}A^{DF} -  n_e^{T}v^{T}(T_e)A^{DF}\frac{R(T_e)}{R^{DF}}= -\int_{T_e^T}^{T_e^{DF}}S_{rec}(T_e)A^{DF}\frac{R(T_e)}{R^{DF}}\frac{\dd x}{\dd T_e}(T_e) \dd T_e
\end{align}
The volumetric particle sinks can be expressed in terms of the effective rates for electron-ion recombination (ACD), taken from the ADAS database, and effective molecular activated recombination (MAR) rates, obtained from AMJUEL\cite{verhaegh_molecules_2_2021}:
\begin{align}
    n_e^{DF}v^{DF} -  n_e^{T}v^{T}(T_e)\frac{R(T_e)}{R^{DF}}= -\int_{T_e^T}^{T_e^{DF}} \left(ACD (n_e,T_e)n_e^2  + MAR (n_e,T_e)n_e D_2(T_e)\right)\frac{\dd x}{\dd T_e}(T_e)\frac{R(T_e)}{R^{DF}} \dd T_e
\end{align}
where the values of $n_e^{DF},v(T_e), D_2(T_e)$ and the temperature gradient $\frac{\dd T}{\dd x}(T_e)$ are taken from the respective interpretative SOLPS simulation.

The equation can then be solved numerically iteratively by starting from the detachment front (assumed as a temperature front at $T_e^{DF}$ = 3 eV) and taking steps towards lower temperatures.
\begin{multline}
    n_e^{i+1} (T_e + \Delta T_e)=  n_e^{DF}\frac{v^{DF}}{v^{i+1} (T_e + \Delta T_e)}\frac{R^{DF}}{R(T_e)} + \\ 
    +  \frac{1}{v^{i+1} (T_e + \Delta T_e)}\frac{R^{DF}}{R(T_e)}\sum_{j = 0} ^{i+1} \left(ACD(n_e^j, T_e^j) \left(n_e^j\right)^2 + MAR (n_e^j,T_e^j)n_e^j D_2(T_e^j)\right)\frac{dx}{dT_e}(T_e^j)\frac{R(T_e^j)}{R^{DF}} \Delta T_e
\end{multline}

The solution of the equation for the SXD NBI heated simulation discussed in section in \ref{sec:SOLPS} are shown in figure \ref{fig19:1D_model}.

To extrapolate the behaviour of the elctron density profile to higher density conditions, the values of $v(T_e,L_T), D_2/n_e(T_e,L_T)$ and $\frac{\dd T}{\dd x}(T_e,L_T)$ are interpolated over a series of SXD SOLPS simulations at different heating powers and fuelling levels as a function of electron temperature and parallel distance from the target ($L_T$). The solution of the continuity equation extrapolated to higher density are shown in figure \ref{fig20:1D_model_extrapolation}.

\section*{References}
\bibliographystyle{iopart-num}
\bibliography{bib}

\end{document}

%% file: packages_commands.tex
 \usepackage{hyperref}
 \usepackage{color}
 \usepackage[utf8]{inputenc}
 \usepackage{bm}
 \usepackage{graphicx, subcaption}
 \usepackage{latexsym}
 \usepackage{pifont}
 \usepackage{amssymb}
 \usepackage{hyperref}
 \usepackage{titlesec}
 \expandafter\let\csname equation*\endcsname\relax
 \expandafter\let\csname endequation*\endcsname\relax
 \usepackage{amsmath}
 \usepackage{physics}
 \usepackage{listings}
 \usepackage{verbatim}
 \usepackage{geometry}
 \usepackage{booktabs}
 \usepackage{enumitem}
 \usepackage[normalem]{ulem} 
 \usepackage{breqn}
 \usepackage[toc,page]{appendix}
 
 \definecolor{red}{rgb}{1.0,0.0,0.0}
 \definecolor{gre}{rgb}{0.0,1.0,0.0}
 \definecolor{blu}{rgb}{0.0,0.0,1.0}

 \definecolor{ora}{rgb}{1.0,0.5,0.0}
 \definecolor{gra}{rgb}{0.0,0.5,1.0}

 \newcommand{\ba}{\begin{abstract}}
 \newcommand{\ea}{\end{abstract}}
 \newcommand{\be}{\begin{equation}}
 \newcommand{\ee}{\end{equation}}

 \newcommand{\rom}[1]{\uppercase\expandafter{\romannumeral #1 \relax}}

%% file: main.bbl
\providecommand{\newblock}{}
\begin{thebibliography}{10}
\expandafter\ifx\csname url\endcsname\relax
  \def\url#1{{\tt #1}}\fi
\expandafter\ifx\csname urlprefix\endcsname\relax\def\urlprefix{URL }\fi
\providecommand{\eprint}[2][]{\url{#2}}

\bibitem{DEMO_lambda}
Wenninger R, Bernert M, Eich T, Fable E, Federici G, Kallenbach A, Loarte A, Lowry C, McDonald D, Neu R, Pütterich T, Schneider P, Sieglin B, Strohmayer G, Reimold F and Wischmeier M 2014 {\em Nuclear Fusion\/} {\bf 54} 114003 \urlprefix\url{https://dx.doi.org/10.1088/0029-5515/54/11/114003}

\bibitem{Wenninger_2014_DEMO_detach}
Wenninger R, Bernert M, Eich T, Fable E, Federici G, Kallenbach A, Loarte A, Lowry C, McDonald D, Neu R, Pütterich T, Schneider P, Sieglin B, Strohmayer G, Reimold F and Wischmeier M 2014 {\em Nuclear Fusion\/} {\bf 54} 114003 \urlprefix\url{https://dx.doi.org/10.1088/0029-5515/54/11/114003}

\bibitem{core_confinement_density_degrad}
(prepared~by GF~Matthews) J~T 1999 {\em Nuclear Fusion\/} {\bf 39} 1687 \urlprefix\url{https://dx.doi.org/10.1088/0029-5515/39/11Y/308}

\bibitem{kool2024demonstrationsuperxdivertorexhaust}
Kool B, Verhaegh K, Derks G~L, Wijkamp T~A, Lonigro N, Doyle R, McArdle G, Vincent C, Lovell J, Federici F, Henderson S~S, Osawa R~T, Brida D, Reimerdes H, van Berkel M, tokamak~exploitation team T~E and the MAST-U~team 2024 First demonstration of super-x divertor exhaust control for transient heat load management in compact fusion reactors (\textit{Preprint} \eprint{2407.07784}) \urlprefix\url{https://arxiv.org/abs/2407.07784}

\bibitem{Kotschenreuther2004ScrapeOL}
Kotschenreuther M~T, Valanju P~M, Wiley J~C, Rognlein T~D, Mahajan S~M and Pekker M 2004 Scrape off layer physics for burning plasmas and innovative divertor solutions \urlprefix\url{https://api.semanticscholar.org/CorpusID:44327246}

\bibitem{Ryutov2007}
Ryutov D 2007 {\em Physics of Plasmas\/} {\bf 14} cited by: 269; All Open Access, Green Open Access \urlprefix\url{https://www.scopus.com/inward/record.uri?eid=2-s2.0-34547380960&doi=10.1063%2f1.2738399&partnerID=40&md5=283f27afc0e1b84c65984891748c176b}

\bibitem{Valanju_2009}
Valanju P~M, Kotschenreuther M, Mahajan S~M and Canik J 2009 {\em Physics of Plasmas\/} {\bf 16} 056110 ISSN 1070-664X (\textit{Preprint} \eprint{https://pubs.aip.org/aip/pop/article-pdf/doi/10.1063/1.3110984/14032310/056110\_1\_online.pdf}) \urlprefix\url{https://doi.org/10.1063/1.3110984}

\bibitem{LaBombard_2015}
LaBombard B, Marmar E, Irby J, Terry J, Vieira R, Wallace G, Whyte D, Wolfe S, Wukitch S, Baek S, Beck W, Bonoli P, Brunner D, Doody J, Ellis R, Ernst D, Fiore C, Freidberg J, Golfinopoulos T, Granetz R, Greenwald M, Hartwig Z, Hubbard A, Hughes J, Hutchinson I, Kessel C, Kotschenreuther M, Leccacorvi R, Lin Y, Lipschultz B, Mahajan S, Minervini J, Mumgaard R, Nygren R, Parker R, Poli F, Porkolab M, Reinke M, Rice J, Rognlien T, Rowan W, Shiraiwa S, Terry D, Theiler C, Titus P, Umansky M, Valanju P, Walk J, White A, Wilson J, Wright G and Zweben S 2015 {\em Nuclear Fusion\/} {\bf 55} 053020 \urlprefix\url{https://dx.doi.org/10.1088/0029-5515/55/5/053020}

\bibitem{Piras_2009}
Piras F, Coda S, Furno I, Moret J~M, Pitts R~A, Sauter O, Tal B, Turri G, Bencze A, Duval B~P, Felici F, Pochelon A and Zucca C 2009 {\em Plasma Physics and Controlled Fusion\/} {\bf 51} 055009 \urlprefix\url{https://dx.doi.org/10.1088/0741-3335/51/5/055009}

\bibitem{maurizio_TCV_2018}
Maurizio R, Tsui C~K, Duval B~P, Reimerdes H, Theiler C, Boedo J, Labit B, Sheikh U, Spolaore M, Team T~T and Team T~E~M 2018 {\em Nuclear Fusion\/} {\bf 59} 016014 ISSN 0029-5515 publisher: IOP Publishing \urlprefix\url{https://dx.doi.org/10.1088/1741-4326/aaee1b}

\bibitem{soukhanovskii_mastu_2022}
Soukhanovskii V~A, Cunningham G, Harrison J~R, Federici F and Ryan P 2022 {\em Nuclear Materials and Energy\/} {\bf 33} 101278 ISSN 2352-1791 \urlprefix\url{https://www.sciencedirect.com/science/article/pii/S2352179122001594}

\bibitem{Verhaegh_2023}
Verhaegh K, Lipschultz B, Harrison J, Osborne N, Williams A, Ryan P, Allcock J, Clark J, Federici F, Kool B, Wijkamp T, Fil A, Moulton D, Myatra O, Thornton A, Bosman T, Bowman C, Cunningham G, Duval B, Henderson S, Scannell R and the MAST Upgrade~team 2022 {\em Nuclear Fusion\/} {\bf 63} 016014 \urlprefix\url{https://dx.doi.org/10.1088/1741-4326/aca10a}

\bibitem{Verhaegh_2023_2}
Verhaegh K, Lipschultz B, Harrison J, Federici F, Moulton D, Lonigro N, Kobussen S, O’Mullane M, Osborne N, Ryan P, Wijkamp T, Kool B, Rose E, Theiler C, Thornton A and the MAST Upgrade~Team 2023 {\em Nuclear Fusion\/} {\bf 63} 126023 \urlprefix\url{https://dx.doi.org/10.1088/1741-4326/acf946}

\bibitem{verhaegh2024improved}
Verhaegh K, Harrison J, Moulton D, Lipschultz B, Lonigro N, Osborne N, Ryan P, Theiler C, Wijkamp T, Brida D, Cowley C, Derks G, Doyle R, Federici F, Kool B, Février O, Hakola A, Henderson S, Reimerdes H, Thornton A, Vianello N, Wischmeier M and Xiang L 2025 {\em Communications Physics\/} {\bf 8} 215 ISSN 2399-3650 publisher: Nature Publishing Group \urlprefix\url{https://www.nature.com/articles/s42005-025-02121-1}

\bibitem{moulton_super-x_2024}
Moulton D, Harrison J~R, Xiang L, Ryan P~J, Kirk A, Verhaegh K, Wijkamp T~A, Federici F, Clark J~G and Lipschultz B 2024 {\em Nuclear Fusion\/} {\bf 64} 076049 ISSN 0029-5515 publisher: IOP Publishing \urlprefix\url{https://dx.doi.org/10.1088/1741-4326/ad4f9c}

\bibitem{MAURIZIO2024101736}
Maurizio R, Leonard A, Yu J, Harrison J, Verhaegh K, Lonigro N and McLean A 2024 {\em Nuclear Materials and Energy\/} {\bf 41} 101736 ISSN 2352-1791 \urlprefix\url{https://www.sciencedirect.com/science/article/pii/S2352179124001595}

\bibitem{Osborne_2024}
Osborne N, Verhaegh K, Bowden M~D, Wijkamp T, Lonigro N, Ryan P, Pawelec E, Lipschultz B, Soukhanovskii V, van~den Biggelaar T and the MAST-U~Team 2023 {\em Plasma Physics and Controlled Fusion\/} {\bf 66} 025008 \urlprefix\url{https://dx.doi.org/10.1088/1361-6587/ad1654}

\bibitem{HENDERSON_reattachment}
Henderson S, Bernert M, Brida D, Derks G, Elmore S, Federici F, Harrison J, Kirk A, Kool B, Lonigro N, Lovell J, Moulton D, Reimerdes H, Ryan P, Stobbs J, Verhaegh K, {van den Doel} T, Wijkamp T and Bardsley O 2024 {\em Nuclear Materials and Energy\/} {\bf 41} 101765 ISSN 2352-1791 \urlprefix\url{https://www.sciencedirect.com/science/article/pii/S2352179124001881}

\bibitem{Harrison_2024}
Harrison J, Aboutaleb A, Ahmed S, Aljunid M, Allan S, Anand H, Andrew Y, Appel L, Ash A, Ashton J, Bachmann O, Barnes M, Barrett B, Baver D, Beckett D, Bennett J, Berkery J, Bernert M, Boeglin W, Bowman C, Bradley J, Brida D, Browning P, Brunetti D, Bryant P, Bryant J, Buchanan J, Bulmer N, Carruthers A, Cecconello M, Chen Z, Clark J, Cowley C, Coy M, Crocker N, Cunningham G, Cziegler I, Assuncao T~D, Damizia Y, Davies P, Day I, Derks G, Dixon S, Doyle R, Dreval M, Dunne M, Duval B, Eagles T, Edmond J, El-Haroun H, Elmore S, Enters Y, Faitsch M, Federici F, Fedorczak N, Felici F, Field A, Fitzgerald M, Fitzgerald I, Fitzpatrick R, Frassinetti L, Fuller W, Gahle D, Galdon-Quiroga J, Garzotti L, Gee S, Gheorghiu T, Gibson S, Gibson K, Giroud C, Greenhouse D, Hall-Chen V, Ham C, Harrison R, Henderson S, Hickling C, Hnat B, Howlett L, Hughes J, Hussain R, Imada K, Jacquet P, Jepson P, Kandan B, Katramados I, Kazakov Y, King D, King R, Kirk A, Knolker M, Kochan M, Kogan L, Kool B, Kotschenreuther M, Lees M,
  Leonard A, Liddiard G, Lipschultz B, Liu Y, Lomanowski B, Lonigro N, Lore J, Lovell J, Mahajan S, Maiden F, Man-Friel C, Mansfield F, Marsden S, Martin R, Mazzi S, McAdams R, McArdle G, McClements K, McClenaghan J, McConville D, McKay K, McKnight C, McKnight P, McLean A, McMillan B, McShee A, Measures J, Mehay N, Michael C, Militello F, Morbey D, Mordijck S, Moulton D, Myatra O, Nelson A, Nicassio M, O’Mullane M, Oliver H, Ollus P, Osborne T, Osborne N, Parr E, Parry B, Patel B, Payne D, Paz-Soldan C, Phelps A, Piron L, Piron C, Prechel G, Price M, Pritchard B, Proudfoot R, Reimerdes H, Rhodes T, Richardson P, Riquezes J, Rivero-Rodriguez J, Roach C, Robson M, Ronald K, Rose E, Ryan P, Ryan D, Saarelma S, Sabbagh S, Sarwar R, Saunders P, Sauter O, Scannell R, Schuett T, Seath R, Sharma R, Shi P, Sieglin B, Simmonds M, Smith J, Smith A, Soukhanovskii V~A, Speirs D, Staebler G, Stephen R, Stevenson P, Stobbs J, Stott M, Stroud C, Tame C, Theiler C, Thomas-Davies N, Thornton A, Tobin M, Vallar M, Vann R,
  Velarde L, Verhaegh K, Viezzer E, Vincent C, Voss G, Warr M, Wehner W, Wiesen S, Wijkamp T, Wilkins D, Williams T, Wilson T, Wilson H, Wong H, Wood M and Zamkovska V 2024 {\em Nuclear Fusion\/} {\bf 64} 112017 \urlprefix\url{https://dx.doi.org/10.1088/1741-4326/ad6011}

\bibitem{FEDERICI2025101940}
Federici F, Reinke M~L, Lipschultz B, Lovell J~J, Verhaegh K, Lonigro N, Cowley C, Kryjak M, Ryan P, Thornton A~J, Harrison J~R, Peterson B~J, Lomanowski B, Lore J~D and Damizia Y 2025 {\em Nuclear Materials and Energy\/} {\bf 43} 101940 ISSN 2352-1791 \urlprefix\url{https://www.sciencedirect.com/science/article/pii/S2352179125000821}

\bibitem{Theiler_2017}
Theiler C, Lipschultz B, Harrison J, Labit B, Reimerdes H, Tsui C, Vijvers W, Boedo J~A, Duval B, Elmore S, Innocente P, Kruezi U, Lunt T, Maurizio R, Nespoli F, Sheikh U, Thornton A, van Limpt S, Verhaegh K, Vianello N, the TCV~team and the EUROfusion MST1~team 2017 {\em Nuclear Fusion\/} {\bf 57} 072008 \urlprefix\url{https://dx.doi.org/10.1088/1741-4326/aa5fb7}

\bibitem{Theiler_2024}
Theiler C, Fevrier O, Reimerdes H, Thornton A, Baquero-Ruiz M, Berner M, Blanchard P, Brida D, Colandrea C, Oliveira H~D, Dunne M, Duval B~P, Fasoli A, Fil A, Frassinetti L, Galassi D, Gorno S, Harrison J, Henderson S, Komm M, Labit B, Linehan B, Lipschultz B, Martinelli L, Offeddu N, Perek A, Raj H, Sheikh U, Sun G, Tsui C~K, Vincent B, Wensing M, Wuethrich C, Team T~T and Team T~E~M (2021) {\em Proc. of the 28th IAEA Fusion Energy Conference, virtual event\/}

\bibitem{Carpita_2024}
Carpita M, Février O, Reimerdes H, Theiler C, Duval B, Colandrea C, Durr-Legoupil-Nicoud G, Galassi D, Gorno S, Huett E, Loizu J, Martinelli L, Perek A, Simons L, Sun G, Tonello E, Wüthrich C and the TCV~Team 2024 {\em Nuclear Fusion\/} {\bf 64} 046019 \urlprefix\url{https://dx.doi.org/10.1088/1741-4326/ad2a2a}

\bibitem{fil_TCV_2020}
Fil A, Lipschultz B, Moulton D, Dudson B~D, Février O, Myatra O, Theiler C, Verhaegh K, Wensing M and {and} 2020 {\em Plasma Physics and Controlled Fusion\/} {\bf 62} 035008 ISSN 0741-3335 publisher: IOP Publishing \urlprefix\url{https://doi.org/10.1088/1361-6587/ab69bb}

\bibitem{Doyle_CIS}
Doyle R~S, Lonigro N, Allcock J~S, Silburn S~A, Turner M~M, Feng X and Leggate H 2024 {\em Review of Scientific Instruments\/} {\bf 95} 053505 ISSN 0034-6748 (\textit{Preprint} \eprint{https://pubs.aip.org/aip/rsi/article-pdf/doi/10.1063/5.0205584/19947615/053505\_1\_5.0205584.pdf}) \urlprefix\url{https://doi.org/10.1063/5.0205584}

\bibitem{lonigro_CIS}
Lonigro N, Doyle R~S, Allcock J~S, Lipschultz B, Verhaegh K, Bowman C, Brida D, Harrison J, Myatra O, Silburn S, Theiler C, Wijkamp T~A, Team M~U and the EUROfusion Tokamak Exploitation~Team 2025 {\em Plasma Physics and Controlled Fusion\/} {\bf 67} 035003 \urlprefix\url{https://dx.doi.org/10.1088/1361-6587/adad97}

\bibitem{LFS_H_mode_access}
Field A~R, Carolan P~G, Conway N~J, Counsell G~F, Cunningham G, Helander P, Meyer H, Taylor D, Tournianski M~R, Walsh M~J and the MAST~team 2004 {\em Plasma Physics and Controlled Fusion\/} {\bf 46} 981 \urlprefix\url{https://dx.doi.org/10.1088/0741-3335/46/6/005}

\bibitem{Loarte_2001_review}
Loarte A 2001 {\em Plasma Physics and Controlled Fusion\/} {\bf 43} R183 \urlprefix\url{https://dx.doi.org/10.1088/0741-3335/43/6/201}

\bibitem{wijkamp_MWI_2023}
Wijkamp T~A, Allcock J~S, Feng X, Kool B, Lipschultz B, Verhaegh K, Duval B~P, Harrison J~R, Kogan L, Lonigro N, Perek A, Ryan P, Sharples R~M, Classen I~G~J, Jaspers R~J~E and team t~M~U 2023 {\em Nuclear Fusion\/} {\bf 63} 056003 ISSN 0029-5515 publisher: IOP Publishing \urlprefix\url{https://dx.doi.org/10.1088/1741-4326/acc191}

\bibitem{Allcock_CIS_density}
Allcock J~S, Silburn S~A, Sharples R~M, Harrison J~R, Conway N~J and Vernimmen J~W~M 2021 {\em Review of Scientific Instruments\/} {\bf 92} ISSN 0034-6748 073506 (\textit{Preprint} \eprint{https://pubs.aip.org/aip/rsi/article-pdf/doi/10.1063/5.0050704/16017405/073506\_1\_online.pdf}) \urlprefix\url{https://doi.org/10.1063/5.0050704}

\bibitem{feng_development_2021}
Feng X, Calcines A, Sharples R~M, Lipschultz B, Perek A, Vijvers W~A~J, Harrison J~R, Allcock J~S, Andrebe Y, Duval B~P, Mumgaard R~T, {MAST-U Team} and {EUROfusion MST1 Team} 2021 {\em Review of Scientific Instruments\/} {\bf 92} 063510 ISSN 0034-6748, 1089-7623 \urlprefix\url{https://aip.scitation.org/doi/10.1063/5.0043533}

\bibitem{verhaegh2024nbi}
Verhaegh K, Harrison J~R, Lipschultz B, Lonigro N, Kobussen S, Moulton D, Osborne N, Ryan P, Theiler C, Wijkamp T, Brida D, Derks G, Doyle R, Federici F, Hakola A, Henderson S, Kool B, Newton S, Osawa R, Pope X, Reimerdes H, Vianello N and Wischmeier M 2024 Investigations of atomic \& molecular processes of nbi-heated discharges in the mast upgrade super-x divertor with implications for reactors (\textit{Preprint} \eprint{2311.08580})

\bibitem{Lipschultz_2016}
Lipschultz B, Parra F~I and Hutchinson I~H 2016 {\em Nuclear Fusion\/} {\bf 56} 056007 \urlprefix\url{https://dx.doi.org/10.1088/0029-5515/56/5/056007}

\bibitem{Cowley_2022}
Cowley C, Lipschultz B, Moulton D and Dudson B 2022 {\em Nuclear Fusion\/} {\bf 62} 086046 \urlprefix\url{https://dx.doi.org/10.1088/1741-4326/ac7a4c}

\bibitem{myatra_solps-iter_2023}
Myatra O, Moulton D, Dudson B, Lipschultz B, Newton S, Verhaegh K and Fil A 2023 {\em Nuclear Fusion\/} {\bf 63} 076030 ISSN 0029-5515 publisher: IOP Publishing \urlprefix\url{https://dx.doi.org/10.1088/1741-4326/acd9da}

\bibitem{lomanowski_Stark_fit_2015}
Lomanowski B~A, Meigs A~G, Sharples R~M, Stamp M and and C~G 2015 {\em Nuclear Fusion\/} {\bf 55} 123028 ISSN 0029-5515 publisher: IOP Publishing \urlprefix\url{https://doi.org/10.1088/0029-5515/55/12/123028}

\bibitem{VERHAEGH_TCV}
Verhaegh K, Lipschultz B, Duval B, Harrison J, Reimerdes H, Theiler C, Labit B, Maurizio R, Marini C, Nespoli F, Sheikh U, Tsui C, Vianello N and Vijvers W 2017 {\em Nuclear Materials and Energy\/} {\bf 12} 1112--1117 ISSN 2352-1791 proceedings of the 22nd International Conference on Plasma Surface Interactions 2016, 22nd PSI \urlprefix\url{https://www.sciencedirect.com/science/article/pii/S2352179116301740}

\bibitem{KARHUNEN_JET}
Karhunen J, Lomanowski B, Solokha V, Aleiferis S, Carvalho P, Groth M, Kumpulainen H, Lawson K, Meigs A and Shaw A 2020 {\em Nuclear Materials and Energy\/} {\bf 25} 100831 ISSN 2352-1791 \urlprefix\url{https://www.sciencedirect.com/science/article/pii/S2352179120301022}

\bibitem{LIPSCHULTZ_recomb}
Lipschultz B, Terry J, Boswell C, Krasheninnikov S, LaBombard B and Pappas D 1999 {\em Journal of Nuclear Materials\/} {\bf 266-269} 370--375 ISSN 0022-3115 \urlprefix\url{https://www.sciencedirect.com/science/article/pii/S0022311598005340}

\bibitem{Stangeby_2018}
Stangeby P~C 2018 {\em Plasma Physics and Controlled Fusion\/} {\bf 60} 044022 \urlprefix\url{https://dx.doi.org/10.1088/1361-6587/aaacf6}

\bibitem{Verhaegh_CX}
Verhaegh K, Williams A, Moulton D, Lipschultz B, Duval B, Février O, Fil A, Harrison J, Osborne N, Reimerdes H, Theiler C, the TCV~Team and the EUROfusion MST1~Team 2023 {\em Nuclear Fusion\/} {\bf 63} 076015 \urlprefix\url{https://dx.doi.org/10.1088/1741-4326/acd394}

\bibitem{osborne2024_NBI}
Osborne N, Verhaegh K, Moulton D, Reimerdes H, Ryan P, Lonigro N, Mijin S, Osawa R, Murray K, Kobussen S, Damizia Y, Perek A, Theiler C, Ducker R and Mykytchuk D 2024 A novel understanding of the role of plasma-molecular kinetics on divertor power exhaust (\textit{Preprint} \eprint{2410.14403}) \urlprefix\url{https://arxiv.org/abs/2410.14403}

\bibitem{verhaegh_molecules_2_2021}
Verhaegh K, Lipschultz B, Harrison J~R, Duval B~P, Fil A, Wensing M, Bowman C, Gahle D~S, Kukushkin A, Moulton D, Perek A, Pshenov A, Federici F, Février O, Myatra O, Smolders A, Theiler C, Team t~T and Team t~E~M 2021 {\em Nuclear Fusion\/} {\bf 61} 106014 ISSN 0029-5515 publisher: IOP Publishing \urlprefix\url{https://doi.org/10.1088/1741-4326/ac1dc5}

\end{thebibliography}
